\documentstyle[12pt,psfig]{article}

\def\3{\ss}                                                   
\def\lsim{\raise.25ex \hbox{ $<$ \kern-1.1em \lower1ex \hbox {$\sim$ }}}
\def\gsim{\raise.25ex \hbox{ $>$ \kern-1.1em \lower1ex \hbox {$\sim$ }}}

\newcommand{\ppbar}{\mbox{$\mathrm{p}\bar{\mathrm{p}}$}}

\newcommand{\qqbar}{\mbox{$\mathrm{q}\bar{\mathrm{q}}$}}

\newcommand{\mev}{{\rm Me}\kern-1.pt{\rm V}}
\newcommand{\gev}{{\rm Ge}\kern-1.pt{\rm V}}
\newcommand{\gevsq}{\mbox{$\mathrm{{\rm Ge}\kern-1.pt{\rm V}}^2$}}
\newcommand{\gevmsq}{\mbox{$\mathrm{{\rm Ge}\kern-1.pt{\rm V}}^{-2}$}}

\begin{document}  

\title {\begin{flushright}{\large DESY--98--107\\}  \end{flushright}
\vspace{0.1cm}
\begin{flushright}{\large hep--ex/9808020\\}  \end{flushright}
\vspace{1cm}
\LARGE \bf  Exclusive Electroproduction of $\rho^0$ and $J/\psi$ Mesons at HERA\\
       \vspace{1cm}}    
\author{\bf ZEUS Collaboration\\ \vspace{2cm}}
\date{ }
\maketitle
\begin{abstract}
\noindent
Exclusive  production of $\rho^0$ and $J/\psi$ mesons 
in   e$^+$p collisions  has been  studied with the ZEUS detector
in the kinematic range 
$0.25 < Q^2 < 50$~{\gevsq}, $20 < W < 167$~{\gev} for the $\rho^0$ data 
and $2 < Q^2 < 40$~{\gevsq}, $50 < W < 150$~{\gev} for the $J/\psi$ data.
Cross sections for  exclusive $\rho^0$ and $J/\psi$ 
production have been  measured as a function  of 
$Q^2$, $W$ and  $t$. The spin-density matrix elements 
$r^{04}_{00}$,  $r^1_{1-1}$ and Re$\:r^{5}_{10}$
have been  determined for  exclusive $\rho^0$ production
as well as $r^{04}_{00}$ and   $r^{04}_{1-1}$
for  exclusive $J/\psi$ production.
The results are discussed in the context of theoretical
models invoking soft and hard phenomena.
\end{abstract}
\pagestyle{plain}

\thispagestyle{empty}

\clearpage
\pagenumbering{Roman} 

%
%
%
%
                                                   %
\begin{center}                                                                                     
{                      \Large  The ZEUS Collaboration              }                               
\end{center}                                                                                       
  J.~Breitweg,                                                                                     
  S.~Chekanov,                                                                                     
  M.~Derrick,                                                                                      
  D.~Krakauer,                                                                                     
  S.~Magill,                                                                                       
  D.~Mikunas,                                                                                      
  B.~Musgrave,                                                                                     
  J.~Repond,                                                                                       
  R.~Stanek,                                                                                       
  R.L.~Talaga,                                                                                     
  R.~Yoshida,                                                                                      
  H.~Zhang  \\                                                                                     
 {\it Argonne National Laboratory, Argonne, IL, USA}~$^{p}$
% ...
\newpage                                                                                           
% ...
$^{  18}$ now at ATLAS Collaboration, Univ. of Munich \\                                           
% ...
$^{  31}$ partially supported by the Foundation for German-Russian Collaboration                   
DFG-RFBR \\ \hspace*{3.5mm} (grant no. 436 RUS 113/248/3 and no. 436 RUS 113/248/2)\\              
% ...
                                                           %
                                                           %
\newpage   
                                                           %
                                                           %
\begin{tabular}[h]{rp{14cm}}                                                                       
$^{a}$ &  supported by the Natural Sciences and Engineering Research                               
          Council of Canada (NSERC)  \\                                                            
$^{b}$ &  supported by the FCAR of Qu\'ebec, Canada  \\                                            
$^{c}$ &  supported by the German Federal Ministry for Education and                               
          Science, Research and Technology (BMBF), under contract                                  
          numbers 057BN19P, 057FR19P, 057HH19P, 057HH29P \\                                        
$^{d}$ &  supported by the MINERVA Gesellschaft f\"ur Forschung GmbH,                              
          the German Israeli Foundation, the U.S.-Israel Binational                                
          Science Foundation, and by the Israel Ministry of Science \\                             
$^{e}$ &  supported by the German-Israeli Foundation, the Israel Science                           
          Foundation, the U.S.-Israel Binational Science Foundation, and by                        
          the Israel Ministry of Science \\                                                        
$^{f}$ &  supported by the Italian National Institute for Nuclear Physics                          
          (INFN) \\                                                                                
$^{g}$ &  supported by the Japanese Ministry of Education, Science and                             
          Culture (the Monbusho) and its grants for Scientific Research \\                         
$^{h}$ &  supported by the Korean Ministry of Education and Korea Science                          
          and Engineering Foundation  \\                                                           
$^{i}$ &  supported by the Netherlands Foundation for Research on                                  
          Matter (FOM) \\                                                                          
$^{j}$ &  supported by the Polish State Committee for Scientific                                   
          Research, grant No.~115/E-343/SPUB/P03/002/97, 2P03B10512,                               
          2P03B10612, 2P03B14212, 2P03B10412 \\                                                    
$^{k}$ &  supported by the Polish State Committee for Scientific                                   
          Research (grant No. 2P03B08614) and Foundation for                                       
          Polish-German Collaboration  \\                                                          
$^{l}$ &  partially supported by the German Federal Ministry for                                   
          Education and Science, Research and Technology (BMBF)  \\                                
$^{m}$ &  supported by the Fund for Fundamental Research of Russian Ministry                       
          for Science and Edu\-cation and by the German Federal Ministry for                       
          Education and Science, Research and Technology (BMBF) \\                                 
$^{n}$ &  supported by the Spanish Ministry of Education                                           
          and Science through funds provided by CICYT \\                                           
$^{o}$ &  supported by the Particle Physics and                                                    
          Astronomy Research Council \\                                                            
$^{p}$ &  supported by the US Department of Energy \\                                              
$^{q}$ &  supported by the US National Science Foundation \\                                       
\end{tabular}                                                                                      
                                                           %
                                                           %
\newpage
\setcounter{page}{1}
\pagenumbering{arabic}
\section{Introduction}

We report  measurements of  exclusive   electroproduction of $\rho^0$ 
and $J/\psi$ mesons  at electron-proton centre-of-mass energy of 
$\sqrt{s}=300$~{\gev} using
the ZEUS detector  at HERA. The reaction  ${\rm e p} \rightarrow {\rm e V p}$,
where $V$ stands for a vector
meson ($\rho^0$, $\phi$, $J/\psi$),
is a rich source of information on   soft  and hard diffractive processes 
as well as on the hadronic properties of the virtual 
photon~\cite{ref:crittenden}. 

\begin{figure}
\centerline{
\psfig
{figure=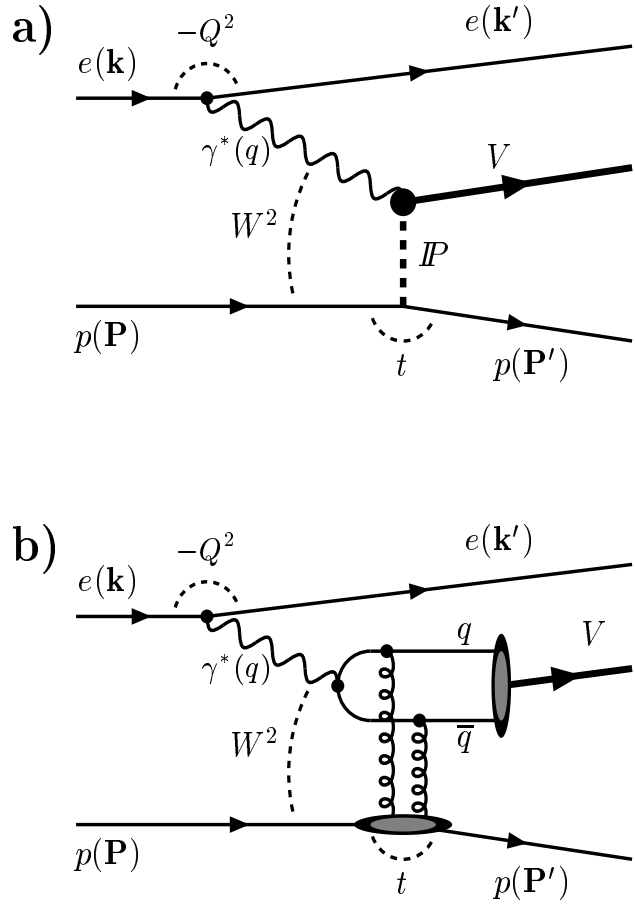,height=12.0cm,width=8.0cm}
}
\caption{
\label{fig:graph}   
Diagrams illustrating the exclusive electroproduction of vector mesons
a)~via pomeron exchange,  and b)~via exchange of a gluon pair.
}  
\end{figure}

Exclusive photoproduction of light vector mesons 
($\rho^0$, $\omega$ and $\phi$)  has been studied in a wide
range of the photon-proton centre-of-mass energy $W$ both in
fixed target experiments~\cite{ref:bauer} 
and at HERA~\cite{ref:rhoZEUS93,ref:jetphi,ref:heravm}.
For $W\gsim 10$~{\gev} these reactions display features characteristic
of a soft diffractive process: $s$-channel helicity conservation,
cross sections  rising weakly with $W$ and a steep exponential
$t$ dependence, where $t$ is the squared four-momentum transfer
at the proton vertex.  Such
processes are well described within the framework of 
Regge phenomenology~\cite{ref:regge} and the Vector-Meson Dominance model 
(VMD)~\cite{ref:vdm},  where exclusive vector-meson (VM)  
production at high energies
is assumed to proceed  via  the exchange of a Pomeron trajectory
as shown in Fig.~\ref{fig:graph}a.  
In this approach, the $W$ and $t$ dependences  are coupled:
\begin{equation}
\frac{{\rm d}\sigma^{\gamma {\rm p}}}{{\rm d}|t|} \propto e^{-b_0|t|}\left(\frac{W^2}{W_0^2}\right)^{
2(\alpha(t)-1)}, 
\end{equation}
\noindent 
where $\alpha(t) = \alpha(0) + \alpha^{\prime}t$, while $b_0$ and $W_0$ are
process-dependent constants.
Fits to hadron-hadron scattering
data and photoproduction data give $\alpha(0)= 1.08$
and $\alpha^{\prime} =0.25$~{\gevsq}~\cite{ref:dl}.  
The slope of the $t$ distribution depends on the 
energy as $b = b_0 + 2\alpha^{\prime} \ln (W^2/W_0^2)$ (often
referred to as ``shrinkage''),
while the effective power of a $W^{\delta}$ dependence of the cross 
section (after integrating over
$t$) is $\delta \simeq 4(\alpha(0) - 1 - \alpha^{\prime}/b)$.
Typically, values of $b\simeq 10$~{\gevmsq} are found in the 
photoproduction of light VMs, leading to $\delta\simeq 0.22$, 
in agreement with measurements.  
This  approach fails to describe the 
recently measured energy dependence of the cross section for 
elastic $J/\psi$ photoproduction at HERA~\cite{ref:psi,ref:zeuselpsi}.  
The measured
slope of the $|t|$ distribution, $b_{J/\psi}\simeq 5$~{\gevmsq},
leads to a prediction of $\delta\simeq 0.14$, in contrast to
the measured value of $\delta\simeq 0.9$.

Exclusive VM electroproduction at high values of $Q^2$
has been studied in  fixed target
experiments~\cite{ref:disvm1,ref:nmc,ref:e665} 
and at HERA~\cite{ref:disvm2,ref:h1dis}. 
The  measurements 
indicate that  the rise of the cross section with $W$  is 
stronger than that expected from Regge theory
although there are large uncertainties in the experimental data\cite{ref:disvm2}.
The $Q^2$ dependence of the cross section
can be described  by  $Q^{-2n}$  with $2\lsim n \lsim 2.5$ and
the $|t|$ dependence has a slope $b$ between 4 and  8~{\gevmsq}.
The vector mesons are found to be produced predominantly 
in the helicity 0 state,  whereas in photoproduction  the production
is mainly in the helicity $\pm$1 states.

In models based on perturbative QCD (pQCD), ${\gamma^{\star} {\rm p} \rightarrow {\rm V p}}$ scattering  is viewed
as a  sequence of events  separated in time in the proton rest 
frame~\cite{ref:bfgms}, as depicted in Fig.~\ref{fig:graph}b.  
The steps are: the photon fluctuates into a {\qqbar} state;
the {\qqbar}  pair scatters on the proton target; and,
the scattered {\qqbar}  pair turns into a vector meson.
The interaction  of  the {\qqbar} pair with the proton  
is mediated in leading order 
by the exchange of two gluons in a colour singlet state.  
In this framework, the cross section is proportional to the square of the
gluon density in the proton.   The scale $\mu^2$, at which $\alpha_{\rm s}$ and the 
gluon density are evaluated, can depend on the mass of the vector meson
$M_{\rm V}$, on $Q^2$ and on $t$.  For $J/\psi$ production the scale
$\mu^2 = [Q^2+M_{J/\psi}^2+|t|]/4$~\cite{ref:ryskin}
has been proposed.
In photoproduction at small $|t|$, the scale would therefore have
the value $\mu^2=2.4$~{\gevsq}.  At this scale, the gluon density
at small $x$ rises as  $xg(x) \propto x^{-0.2}$~\cite{ref:gluon}, yielding a $W$
dependence of the  cross section $\sigma^{\gamma {\rm p}} \propto W^{0.8}$,
 significantly steeper than expected from VMD and Regge
phenomenology. The calculation has been extended~\cite{ref:ryslla}  
and compared to HERA  data.   It was found that this process is
indeed  sensitive to the form of the gluon density in the
proton.
Other pQCD calculations have been performed within the
framework of the  Colour Dipole Model (CDM)~\cite{ref:cdm} where
diffraction is viewed as the elastic scattering of a colour dipole
of definite size off the target proton.

At large $Q^2$, the cross section is predicted to be dominated by 
longitudinally polarised virtual photons 
scattering into vector mesons of helicity state 0~\cite{ref:bfgms,ref:dl_vm,ref:ginzburg,ref:collins}. 
The cross section, calculated in leading
$\alpha_{\rm s} \ln\frac{Q^2}{\Lambda^2}\ln\frac{1}{x}$ 
approximation~\cite{ref:bfgms} for vector mesons composed of
light flavours, is
\begin{equation}\label{eq:cross}
\left.\frac{{\rm d}\sigma_{\rm L}^{\gamma^{\star}{\rm p}}}{{\rm d}|t|}\right|_{t=0} = \frac{A}{Q^6}\alpha_{\rm s}^2(Q^2)
 \left| \left( 1 + \frac{{\rm i}\pi}{2} \frac{{\rm d}}{{\rm d}\ln x} \right) 
 xg(x,Q^2)\right|^2,
\end{equation}
where $A$ is a constant which depends on the VM wavefunction.  
Here we discuss some of the expectations for  the cross sections:

\begin{itemize}
\item
The cross section contains a $1/Q^6$ factor.  However,
the $Q^2$ dependences of $\alpha_{\rm s}$ and the gluon density 
also need to be taken into account. 
The effective
$Q^2$ dependence using the CTEQ3L gluon density function~\cite{ref:CTEQ3} and the 
leading-order form for $\alpha_{\rm s}$ is approximately 
${\rm d}\sigma_{\rm L}^{\gamma^{\star}{\rm p}}/{\rm d}|t| \propto 1/Q^5$, with a weak $x$ dependence.
The calculation presented in~\cite{ref:bfgms} has been redone
in leading $\alpha_{\rm s} \ln (Q^2/\Lambda^2)$ 
approximation~\cite{ref:fks}.  
In this work, among other improvements, the
Fermi motion of the quarks in the vector meson has been
considered.  The net effect is to reduce the steepness
of the $Q^2$ dependence.  Precise measurements 
could therefore yield information on the
wavefunctions of the vector mesons.

\item
In the pQCD calculations,  the $t$ and $W$ dependences are not coupled, 
so that no shrinkage is expected.  A lack of
shrinkage, along with a
steep $W$ dependence, indicates that the reaction is
predominantly driven by perturbative processes. 
In such processes, the transverse size of the {\qqbar} pair is small, and the slope is determined by the proton size, resulting in a value for $b$ near 5~{\gevmsq}~\cite{ref:fs_dis95}.
\item
The cross section presented in Eq.~\ref{eq:cross} is for 
longitudinally polarised
photons. The authors of ref.~\cite{ref:bfgms} expect that this is the
dominant contribution to the cross section in DIS.
It has been  argued~\cite{ref:ginzburg} that the
region of validity of the pQCD calculations is signalled by the
predominance of VM production in the helicity zero state.
A recent pQCD calculation for $\rho^0$ 
electroproduction~\cite{ref:mrt},
based on the production of light {\qqbar}
pairs and parton--hadron duality, gives 
an estimate of the transverse photon contribution to
the $\gamma^{\star} {\rm p} \rightarrow {\rho^0} {\rm p}$ cross section.  

\item
The interaction should be flavour-independent at sufficiently high  scales.
From the quark charges of the vector mesons and  assuming a flavour-independent production mechanism, the exclusive production cross sections
should be in the proportions $9:1:2:8$ for $\rho^0:\omega:\phi:J/\psi$. This expectation is
badly broken at low $Q^2$. The pQCD predictions change the ratio somewhat due to
wavefunction effects, such that the relative contribution of heavier 
vector mesons is enhanced~\cite{ref:fks} at high $Q^2$.
\end{itemize}

In this paper, we investigate the dependence of $\rho^0$ and
$J/\psi$ production on the variables $W$, $Q^2$, and $t$.  The $\rho^0$
and $J/\psi$ mesons are identified via their decay to two oppositely charged particles. Invariant masses are reconstructed under the assumption of dipion final states for the $\rho^0$ and dimuon final states for the $J/\psi$.
The decay angular
distributions are also measured, and the helicity matrix elements
extracted, yielding a measurement of $R=\sigma_{\rm L}^{\gamma^{\star}{\rm p}} / \sigma_{\rm T}^{\gamma^{\star}{\rm p}}$ 
as a function of
$Q^2$ and $W$.  The results are compared to expectations from Regge
theory as well as from pQCD models. The data are presented for
$\rho^0$ production in the ranges $0.25<Q^2<0.85$~{\gevsq} (BPC $\rho^0$)
and $3<Q^2<50$~{\gevsq} (DIS $\rho^0$).  The production of $J/\psi$ mesons is 
investigated in the  $Q^2$ range  $2<Q^2<40$~{\gevsq} (DIS $J/\psi$).
The data discussed  in this paper correspond to integrated 
luminosities of 6.0 pb$^{-1}$ (DIS $\rho^0$ and $J/\psi$) and 3.8 pb$^{-1}$ 
(BPC $\rho^0$) collected in 1995.


\section{Experiment}
The measurements were   performed  at the DESY {\rm ep}
collider HERA using the ZEUS detector.
In 1995 HERA operated at a proton energy of 820~{\gev} and a positron
energy of 27.5~{\gev}. 
A detailed description of the ZEUS detector
can be found elsewhere~\cite{ref:zeus}.
The main components used in this analysis are described below.
\par\medskip\noindent
The high-resolution uranium-scintillator calorimeter CAL~\cite{ref:cal} 
consists of three parts:  forward  
\footnote{Throughout this paper the standard ZEUS right-handed
coordinate system is used: the $Z$-axis points in the direction of
the proton beam momentum (referred to as the 
forward direction) and the horizontal $X$-axis
points towards the centre of HERA. The nominal interaction
point is at $X=Y=Z=0$.}
(FCAL),
barrel  (BCAL) and  rear (RCAL) calorimeters.
Each part is subdivided transversely into towers which
are segmented longitudinally  into one electromagnetic
section (EMC) and one (RCAL) or two (FCAL, BCAL)
hadronic sections (HAC).
The energy resolution of the calorimeter, determined in a test beam, 
is $\sigma_{E}/E = 0.18/\sqrt{E}$ for electrons 
and $\sigma_{E}/E = 0.35/\sqrt{E}$ for hadrons,
where $E$ is expressed in~{\gev}.
\par\medskip\noindent
Charged-particle tracks  are reconstructed and their momenta determined
using  the central (CTD)~\cite{ref:ctd} and 
rear tracking detectors (RTD)~\cite{ref:zeus}. 
The CTD is a cylindrical drift chamber operated 
in a magnetic field of 1.43 T
produced by a  superconducting solenoid.
It consists of 72 cylindrical layers, organised in 9 superlayers
covering the polar angular region $15^\circ  < \theta < 164^\circ $.
The RTD is a set of planar drift chambers located at the rear of the CTD,
covering the polar angle region $162^\circ  < \theta < 170^\circ $.
\par\medskip\noindent
The positions of  positrons scattered at small angles with respect to the
beam  direction are determined in  the beam pipe calorimeter (BPC)
and the small-angle rear tracking detector (SRTD). 
\par\medskip\noindent
The BPC 
is an electromagnetic sampling calorimeter 
consisting of 2 modules, placed one on each side of the beam pipe, 
294 cm away from the nominal  ep interaction point in the rear region. 
Each module is equipped with  26 tungsten plates 
(roughly 24 radiation lengths), 
separated by layers of scintillator fingers (strips) each 8 mm wide. 
The strips alternate in the 
horizontal and vertical orientation, providing two-dimensional 
position information.
The energy and position resolutions  for electrons in  the BPC, 
measured in a  test beam,   were  found to be
$17\%/\sqrt{E}$ ($E$ in~{\gev}) and
$\simeq 1$ mm, respectively.
\par\medskip\noindent
The~SRTD~is attached to the front face of the RCAL.
It consists of two planes of scintillator strips, 1 cm wide
and 0.5 cm thick, arranged in orthogonal orientations  and read out
via optical fibres and photomultiplier tubes. It covers the region
of $68\times68$ cm$^2$ in $X$ and $Y$ with the  exclusion of a
$10\times20$ cm$^2$ hole at the centre for the beam pipe. 
The SRTD has a  position resolution of 0.3 cm.
\par\medskip\noindent
The luminosity is determined from the rate of the
Bethe-Heitler process ${\rm e}^{+} {\rm p} \rightarrow {\rm e}^{+} \gamma {\rm p}$,
where the high-energy photon is measured with a lead-scintillator 
calorimeter (LUMI) located at $Z= - 107$~m in the HERA tunnel 
downstream of the interaction point in the
positron flight direction~\cite{ref:lumi}.


\section{Kinematics and cross sections}\label{sec:kinem}

We will use the following kinematic variables to describe
exclusive VM production (see Fig.~\ref{fig:graph}):
$k,\:k^{\prime},\:P,\:P^{\prime},\:q$, the
four-momenta of the incident positron, scattered positron,
incident proton, scattered proton  and virtual photon, respectively;
$Q^2=-q^2=-(k-k^{\prime})^2$, the negative  four-momentum squared of 
the virtual photon; $W^2 = (q+P)^2$, the squared invariant mass  of the 
photon-proton system; $y = (P\cdot q)/(P\cdot k)$, the 
fraction of the positron energy transferred to the photon  
in the  proton rest frame; and,
$t = (P-P^{\prime})^2$,  the squared four-momentum transfer
at the proton vertex. 
\par
\begin{figure}
\centerline{
\psfig
{figure=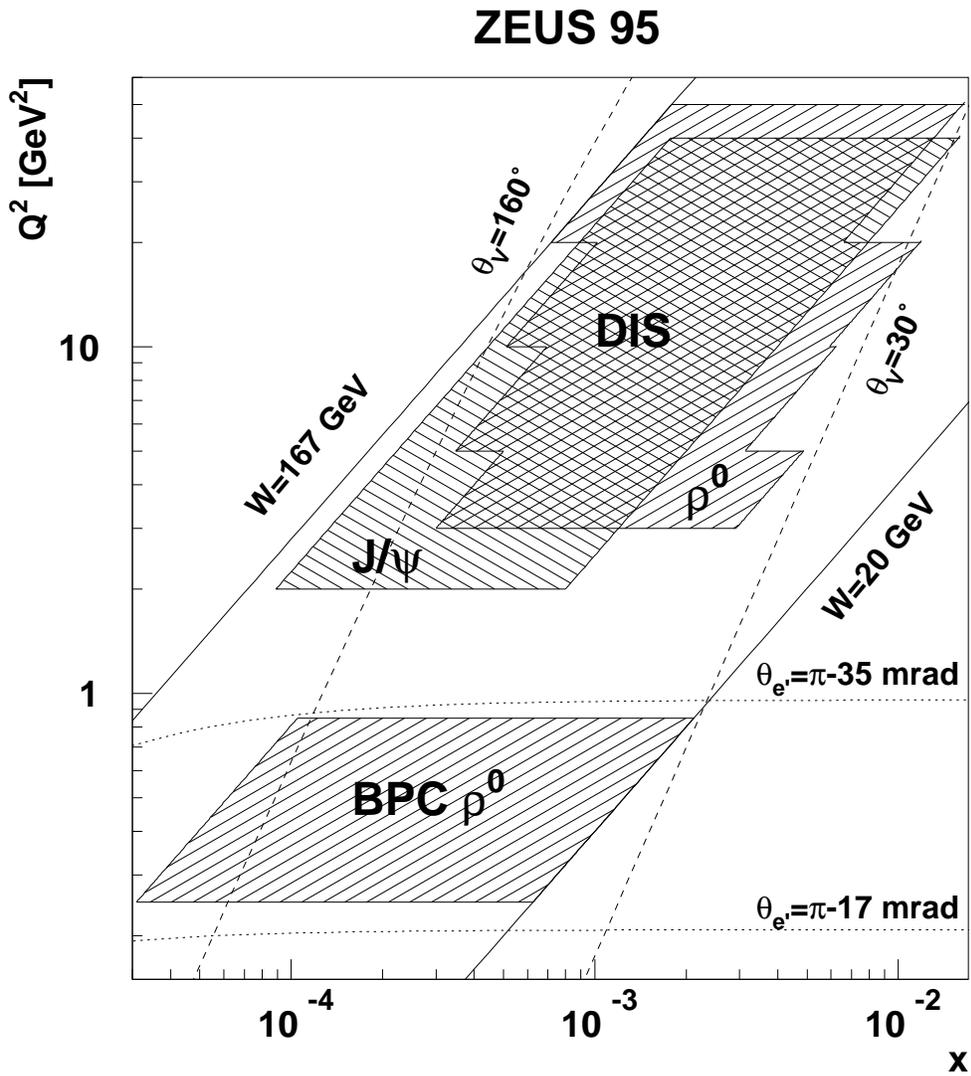,width=\textwidth}
}
\caption{
\label{fig:kinematics}   
The kinematic regions covered by the 
$\rho^0$ and $J/\psi$ data samples used for this analysis.
Line types: solid -- constant W; dashed -- constant polar scattering angle of the vector meson ($\theta_{\rm V}$);  
and, dotted -- constant positron scattering angle ($\theta_{{\rm e}^{\prime}}$).}  
\end{figure}

\medskip\noindent
The kinematic variables were  reconstructed 
with the so-called ``constrained'' method, which involves 
the momenta of the decay  particles measured in  the CTD and
the polar and azimuthal angles of the scattered positron
in the  BPC ($\rho^0$ at low $Q^2$ only) 
or in the main ZEUS calorimeter and the SRTD (all high-$Q^2$ events).
Neglecting the transverse momentum of the outgoing proton
with respect to its incoming momentum,
the energy of the scattered positron can be expressed as
\begin{equation}
E_{{\rm e}^{\prime}} \simeq [2E_{\rm e} -(E_{\rm V} - p^Z_{\rm V})]/(1-\cos{\theta_{{\rm e}^{\prime}}}),
\end{equation}
where $E_{\rm e}$ is the energy of the incident positron,
$E_{\rm V}$ and $p^Z_{\rm V}$ are the energy and longitudinal momentum of the VM,
and $\theta_{{\rm e}^{\prime}}$ is the polar angle of the scattered positron.
The values of $Q^2$  and $t$ were  calculated according to
\begin{equation}
Q^2 = 2E_{{\rm e}^{\prime}} E_{{\rm e}} (1 + \cos{\theta_{{\rm e}^{\prime}}}),
\end{equation}
\begin{equation}
|t| = (p^X_{{\rm e}^{\prime}}+p^X_{\rm V})^{2}+(p^Y_{{\rm e}^{\prime}}+p^Y_{\rm V})^{2},
\end{equation}
where $p^X_{{\rm e}^{\prime}}$, $p^Y_{{\rm e}^{\prime}}$ and $p^X_{\rm V}$, $p^Y_{\rm V}$
are the $X$ and $Y$ components of the momentum of the scattered positron and
VM.
The variable $y$ was calculated  according to  the expression
\begin{equation}
y = (E_{\rm V}-p^Z_{\rm V})/2E_{\rm e}
\end{equation}
and  Bjorken-$x$  was evaluated  using  the relation
$Q^2=sxy$, where $s$ is the squared ep centre of mass energy.
The kinematic ranges  covered by the data are 
shown in Fig.~\ref{fig:kinematics}.
\par\medskip\noindent
In the Born approximation, the  positron-proton cross section can be 
expressed in terms of the transverse, $\sigma_{\rm T}^{\gamma^{\star}{\rm p}}$,  
and longitudinal, $\sigma_{\rm L}^{\gamma^{\star}{\rm p}}$,
virtual photoproduction cross sections as
\begin{equation}\label{eq:d2epdydq2}
\frac{{\rm d}^{2}\sigma^{ep}}{{\rm d}y{\rm d}Q^{2}} = \Gamma_T(y,Q^{2})
(\sigma_{\rm T}^{\gamma^{\star}{\rm p}}+\epsilon
\sigma_{\rm L}^{\gamma^{\star}{\rm p}}), 
\end{equation}
where $\Gamma_{T}$ is the transverse photon flux 
and $\epsilon$ is the ratio of longitudinal and transverse photon fluxes, given by
$\epsilon=2(1-y)/(1+(1-y)^2)$. 
In the kinematic range of this
analysis, the value of  $\epsilon$ varies from 0.94  to 1.0.
The transverse photon flux is~\cite{ref:hand}
\begin{equation}\label{eq:flux}
\Gamma_{T}=\frac{\alpha}{2\pi}\frac{1+(1-y)^{2}}{yQ^{2}},
\end{equation}
where $\alpha$ is the fine-structure constant.
The virtual photon-proton cross section
$\sigma^{\gamma^{\star}{\rm p}} \equiv \sigma_{\rm T}^{\gamma^{\star}{\rm p}}+ \
\epsilon \sigma_{\rm L}^{\gamma^{\star}{\rm p}}$ can be used to evaluate
the total exclusive cross section,   
$\sigma_{\rm tot}^{\gamma^{\star}{\rm p}} \equiv \sigma_{\rm T}^{\gamma^{\star}{\rm p}}+ \
\sigma_{\rm L}^{\gamma^{\star}{\rm p}}$,  through the relation:
\begin{equation}\label{eq:sigmatot}
\sigma_{\rm tot}^{\gamma^{\star}{\rm p}} = \
\frac{1+R}{1+\epsilon R} \sigma^{\gamma^{\star}{\rm p}},
\end{equation}
where $R=\sigma_{\rm L}^{\gamma^{\star}{\rm p}} / \sigma_{\rm T}^{\gamma^{\star}{\rm p}}$ 
is  the ratio of the cross sections for longitudinal and transverse photons.

At given values of $W$  and  $Q^2$,
the exclusive production and decay  of VMs is
described by  three angles:
$\Phi_{\rm h}$ --  the angle between the VM  production plane and the
positron 
scattering plane;
$\theta_{\rm h}$ and $\phi_{\rm h}$ --  the polar and azimuthal angles 
of the positively charged decay particle in the $s$-channel helicity
frame, in which  the spin quantisation  axis is defined along the  VM  direction in
the
photon-proton centre-of-mass system. 
The angular distribution as a function  of these three angles, 
$W(\cos\theta_{\rm h},\phi_{\rm h},\Phi_{\rm h})$, is described by the 
$\rho^0$ spin-density matrix elements, $\rho_{ik}^{\alpha}$,
where $i,k$=-1,0,1 and  by convention $\alpha$=0,1,2,4,5,6 for 
an unpolarised (or transversely polarised) electron 
beam~\cite{ref:angle}. 
The superscripts denote the decomposition of the
spin-density matrix into contributions from the photon polarisation
states: unpolarised transverse photons (0), linearly polarised
transverse photons (1,2), longitudinally polarised photons (4), and
from the interference of longitudinal and transverse amplitudes
(5,6). For  given values of $W$ and $Q^2$,
the polarisation parameter $\epsilon$ is constant, so that  the contributions
from $\rho^{0}_{ik}$ and $\rho^{4}_{ik}$  cannot be distinguished.
The decay angular distribution can therefore be expressed in terms of
linear combinations of the density matrix elements, $r^{04}_{ik}$
and $r^{\alpha}_{ik}$, as
\begin{eqnarray}
r^{04}_{ik} &=& {\rho^0_{ik} \, + \, \epsilon R \rho^{4}_{ik}\over 1 \,
+ \, \epsilon R}, \\
r^{\alpha}_{ik} &=& \,
\left\{
\begin{array}{ll}
{\displaystyle {\rho^{\alpha}_{ik}\over 1 \, + \, \epsilon R}}, &
{\alpha}=1,2\\*[5mm]
{\displaystyle {\sqrt{R} \; \rho^{\alpha}_{ik}\over 1 \, + \, \epsilon
R}}, & {\alpha}=5,6.
\end{array} \right.
\end{eqnarray}
Under the assumption of $s$-channel helicity conservation (SCHC), the 
angular distribution for the decay of the $\rho^0$ meson 
depends on only two angles, $\theta_{\rm h}$ 
and $\psi_{\rm h}=\phi_{\rm h}-\Phi_{\rm h}$, and is characterised
by three independent parameters, $r^{04}_{00}, r^1_{1-1}$ and Re $r^5_{10}$, as
\begin{eqnarray}\label{eq:angular}
W(\cos{\theta_{\rm h}},\psi_{\rm h})&  = & \frac{3}{4\pi}
[\frac{1}{2}(1-r^{04}_{00})+\frac{1}{2}(3r^{04}_{00}-1) \
\cos^2{\theta_{\rm h}} \nonumber  \\
& & +\;\epsilon r^1_{1-1} \sin^2{\theta_{\rm h}} \cos{2\psi_{\rm h}} \
-2\sqrt{\epsilon(1+\epsilon)} {\rm Re}(r^5_{10}) \sin{2\theta_{\rm h}}
\cos{\psi_{\rm h}}]. 
\end{eqnarray}
The ratio $R=\sigma_{\rm L}^{\gamma^{\star}{\rm p}} / \sigma_{\rm T}^{\gamma^{\star}{\rm p}}$ can be determined from the polar
angle distribution via
\begin{equation}\label{eq:R_ratio}
R = \frac{1}{\epsilon}  \frac{r^{04}_{00}}{1 - r^{04}_{00}}.
\end{equation}
The additional assumption of natural-parity exchange in the 
$t$-channel  reduces the number of independent parameters to two. 
The polar and azimuthal angle distributions are related via   
\begin{equation}\label{eq:NPE}
r^1_{1-1}=\frac{1}{2}(1-r^{04}_{00}), 
\end{equation}
independently of the value of $R$. These relations were
found to  hold for diffractive processes at low
energy~\cite{ref:schc}.

Statistical considerations limited our helicity analysis  
of the  $J/\psi$ sample to the one-dimensional distributions in $\theta_{\rm h}$
and $\phi_{\rm h}$, integrated over $\Phi_{\rm h}$. For the decay to spin-1/2 fermions, the above assumptions of SCHC and natural-parity exchange in the $t$-channel yield distributions sensitive to $r^{04}_{00}$ and $r^{04}_{1-1}$
according to
\begin{equation}\label{eq:jpsi_angular1}
W(\cos{\theta_{\rm h}}) = \frac{3}{8}\left[ 1 + r^{04}_{00} 
+ (1 - 3 r^{04}_{00})\cos^2\theta_{\rm h} \right], 
\end{equation}
\begin{equation}\label{eq:jpsi_angular2}
W(\cos{\phi_{\rm h}}) = \frac{1}{2\pi}\left[ 1 + r^{04}_{1-1}\cos 2\phi_{\rm h}  \right]. 
\end{equation}
A value for $R$ can be extracted from the polar angle distribution 
using Eq.~\ref{eq:R_ratio}.

\section{Event selection}

The online event selection is done with a three-level trigger 
system~\cite{ref:trigger}.
The exclusive reaction ${\rm e p} \rightarrow {\rm e} \rho^0 {\rm p}$  at low $Q^2$ (BPC)
was selected at the first trigger level by the requirement 
of an energy deposit in the BPC of more than 6~{\gev} and at 
least one track candidate in the CTD. 
The DIS $\rho^0$ and $J/\psi$  trigger at the first level 
performed an initial identification of 
the scattered electron in the main calorimeter by looking for isolated 
energy deposits.

At the second trigger  level general timing cuts were applied,
along with a cut on the quantity  $E-p_Z = \sum_i (E_i - p_{Z_i})$,
where $E_i=\sqrt{p_{X,i}^2+p_{Y,i}^2+p_{Z,i}^2}$ denotes the energy
in the $i$-th calorimeter cell.
The latter  cut  rejected background from photoproduction events.
The BPC trigger included a restriction on the number of tracks in the CTD.

At the third trigger level, requirements  specific to the exclusive reaction were imposed. 
These requirements were
similar to those applied offline and  included a vertex cut, a limit on the
maximum number of tracks reconstructed by the CTD and
a restriction on the maximum energy in the inner rings of the FCAL
(rejecting events with proton dissociation).
In the case of the DIS $\rho^0$ and $J/\psi$ events a  positron candidate 
in the CAL was required   and a fiducial cut on the positron  
position close to the rear beam pipe  performed.

In the offline selection of  the exclusive  $\rho^0$  and $J/\psi$ 
candidate  events,  the following further  requirements were imposed:
\begin{itemize}
\item
The energy of the scattered positron was required to be greater than  20~{\gev} if
 measured  in the BPC
and greater than 5 (DIS $\rho^0$) or  8~{\gev} ($J/\psi$)  if measured   in  the uranium calorimeter.
Positron identification  in the latter two analyses used an algorithm 
based on a neural 
network~\cite{ref:neural}. The efficiency was
greater than $96\%$. 
In the case of the BPC, cuts were imposed on the  deposited energy, 
shower width, timing and  the BPC fiducial region.
With these cuts, the probability  that 
a selected particle is a positron exceeds $99\%$~\cite{ref:teresa}.
\item
$E-p_Z\:>\:40$~{\gev}. This cut, applied in both DIS  analyses,  
excluded events  requiring  large radiative 
corrections.
\item
The Z coordinate of the interaction vertex was required to
be within $\pm50$ cm of the nominal interaction point.
\item
In addition to the scattered positron the presence of two oppositely charged 
tracks was required, each associated 
with the  reconstructed vertex, and
each with  pseudo\-rap\-i\-di\-ty\footnote{The pseudorapidity $\eta$
is defined as $\rm{\eta = -ln[tan(\frac{\theta}{2})]}$.}
$\left| \eta \right|$ less than 1.75 and  transverse momentum 
greater than 150~{\mev}. These cuts excluded regions of low 
efficiency and poor momentum resolution in the tracking detectors. 
\item
A match between each of the aforementioned tracks and an energy deposit 
in the calorimeter was required.
Energy deposits not associated with tracks or the positron were required to
be less than than 300~{\mev} (elasticity cut), using a matching
procedure developed for this analysis~\cite{ref:beier}.  
\end{itemize}


In addition, the following  cuts were applied to select 
kinematic  regions of high acceptance.
The BPC $\rho^0$ analysis was limited to the region
$0.25<Q^2<0.85$~{\gevsq} and $20 < W < 90$~{\gev}.
The DIS  $\rho^0$ analysis was restricted to the
kinematic region  $3<Q^2<50$~{\gevsq} and $32 < W < 167$~{\gev}.
For the cross section calculation,  only events in the $\pi^+\pi^-$
mass  interval  $0.6<M_{\pi\pi}<1.2$~{\gev} and with
$|t|<0.6$~{\gevsq} were taken in both $\rho^0$ analyses.
For the $J/\psi$ analysis, cuts of
$2<Q^2<40$~{\gevsq} and  $50 < W < 150$~{\gev}  were applied. Only events
within  the mass interval  $2 < M_{{\rm l}^+{\rm l}^-}  < 4$~{\gev} were accepted,
where $M_{{\rm l}^+{\rm l}^-}$ was calculated using the muon mass
for the $J/\psi$ decay products.

The above selection procedure yielded 5462 events in the BPC $\rho^0$ sample,
3039 events in the DIS $\rho^0$ sample and 213 events in the $J/\psi$ sample.


\section{Acceptance  corrections}
In the BPC $\rho^0$ analysis, a  dedicated Monte Carlo generator 
based on  the JETSET~\cite{ref:jetset} package
was used to evaluate the acceptance and resolution of 
the ZEUS detector. The  simulation of exclusive $\rho^0$ production
was based on the VMD model   and Regge phenomenology. 
Events were generated in the region  
0.15 $<Q^2<$ 1.1~{\gevsq}, and 15 $<W<$ 110~{\gev}.
The effective $Q^2$, $W$ and $t$ dependences   of the cross section
were parameterised as  $\sigma_{\rm tot}^{\gamma^{\star}{\rm p}} \propto 1/(1+Q^2/M_{\rho}^2)^{1.75}$, 
$\sigma_{\rm tot}^{\gamma^{\star}{\rm p}} \propto$ $W^{0.12}$ and  ${\rm d}\sigma^{\rm ep} / {\rm d}|t|={\rm exp}(-b|t|+ct^2)$
($b= 9$~{\gevmsq}, $c = 2$~{\gev}$^{-4}$), respectively.
Decay angular distributions  were generated assuming SCHC.
A sample of events was generated using  HERACLES~\cite{ref:heracles} in order to 
evaluate the magnitude  of radiative corrections in the BPC $\rho^0$ data.
For the selected events they  were found not to exceed 2\% for any 
data point and to be   consistent with zero within statistical errors. 
A 2\% error was thus included in the normalisation uncertainty.

In  the DIS $\rho^0$  analysis,
a dedicated program~\cite{ref:muchor} interfaced to HERACLES~\cite{ref:heracles}
was used to evaluate the acceptance and resolution associated with the
``constrained'' method of reconstruction.
The cross section for  exclusive  $\rho^0$ production was parameterised 
in terms  of $\sigma_{\rm L}^{\gamma^{\star}{\rm p}}$ and $\sigma_{\rm T}^{\gamma^{\star}{\rm p}}$ over the entire  $W$ and $Q^2$
range  covered by the data. 
At high $Q^2$, initial-state radiation (ISR) introduces not only
an overall correction to the cross section 
but also significantly distorts  the distributions of 
certain kinematic variables.
In the  ``constrained''  method,  ISR
leads to migration of events along lines of constant
$W$  towards higher values of $Q^2$. 
Moreover, it leads to additional and biased smearing of the reconstructed
value of $t$. (For a cut on $E-p_Z$ of 40~{\gev},
smearing due to ISR produces  a decrease of the $t$ slope by 5\%).
Final-state radiation does not introduce a significant 
error, as the radiated photon is usually well contained within the
calorimeter cluster of the scattered positron.

In the  DIS $J/\psi$ analysis  the
Monte Carlo program DIPSI~\cite{ref:dipsi}, based on the
model of Ryskin~\cite{ref:ryskin}, was used.
In this model,  the exchanged
photon fluctuates into a $c\bar{c}$ pair which 
subsequently  interacts with a gluon
ladder emitted by the incident proton,
and SCHC is assumed.
The longitudinal and transverse cross sections are related by
$\sigma_{\rm L}^{\gamma^{\star}{\rm p}} = (Q^{2}/M_{J/\psi}^{2})\;\sigma_{\rm T}^{\gamma^{\star}{\rm p}}$, where $M_{J/\psi}$ is the mass
of vector meson $J/\psi$. For the  gluon density in the proton  
the MRSA$^{\prime}$ parameterisation was used~\cite{ref:mrsaprime}. 
Events were  generated  assuming  an exponential
$t$ distribution  ${\rm exp}(-b|t|)$ with $b=5$~{\gevmsq}. 
The same method and tools as in the BPC  $\rho^0$ analysis  were used to
calculate  radiative  corrections.
Their  magnitude  was found   not to exceed
4\% for any  data point and this value was  included in the normalisation 
uncertainty.

Two other generators, PYTHIA and EPSOFT,  were used for determination 
of the background from processes in which the proton dissociates.
For the DIS $\rho^0$ and $J/\psi$ analyses, 
the EPSOFT~\cite{ref:epsoft} Monte Carlo, developed in the framework of 
HERWIG~\cite{ref:herwig}, was used. 
It was assumed that  the cross section for the reaction 
$\gamma^{\star} {\rm p} \rightarrow {\rm V  N}$, where N denotes the hadronic final state originating from the dissociated proton, is of the form
\begin{eqnarray}
\frac{{\rm d}^2\sigma^{\gamma^{\star} {\rm p} \rightarrow {\rm V  N}}}{{\rm d}t{\rm d}M_{\rm N}^2}=\frac{1}{2} 
\frac{{\rm d}\sigma^{\gamma^{\star} {\rm p} \rightarrow {\rm V p}}}{{\rm d}|t|}
\left(\frac{{\rm d}\sigma^{{\rm pp} \rightarrow {\rm pN}}}{{\rm d}t{\rm d}M_{\rm N}^2}/\frac{{\rm d}\sigma^{{\rm pp} \rightarrow {\rm pp}}}{{\rm d}t}\right),
\label{diss_1}  
\end{eqnarray}
\noindent
where the ratio 
$\frac{{\rm d}\sigma^{{\rm pp}\rightarrow{\rm pN}}}{{\rm d}t{\rm d}M_{\rm N}^2}/\frac{{\rm d}\sigma^{{\rm pp} \rightarrow {\rm pp}}}{{\rm d}t}$
is obtained from fits to pp data~\cite{ref:epsoft}.
The PYTHIA generator~\cite{ref:pythia} was  used for the BPC analysis.
A cross section of the form 
$\frac{{\rm d}^2\sigma^{\gamma^{\star} {\rm p} \rightarrow {\rm V  N}}}{{\rm d}t{\rm d}M_{\rm N}^2}\propto e^{-b|t|}F_{\rm sd}(M_{\rm N})/M_{\rm N}^2$ 
is assumed in PYTHIA,
with $b=b_0+2\alpha^{\prime} \ln{(W^2/M_{\rm N}^2)}$, $b_0=2.8$~{\gevmsq} and 
$\alpha^{\prime}=0.25$~{\gevsq},
corresponding  to an effective $b\simeq 5$~{\gevmsq} in the kinematic region
of our results. The function $F_{\rm sd}(M_{\rm N})$ enhances 
the cross section in the 
low mass resonance region and suppresses the production 
of very large masses~\cite{ref:pythia}. A fit to the generated $M_{\rm N}$ spectrum
for $10<M_{\rm N}^2<200$~{\gevsq} with a function of the type 
$1/M_{\rm N}^n$ gives $n=2.2$.
The effect of the functions $F_{\rm sd}(M_{\rm N})$ and $b=b(M_{\rm N})$ 
on the spectrum is thus
consistent with the result $n = 2.24 \pm 0.03$
measured for the diffractive dissociation of the
proton in {\ppbar} collisions~\cite{ref:flab}.

In all three analyses 
the generated events were processed  through the same chain of selection and 
reconstruction procedures  as the data,  accounting for  trigger as well as
detector efficiencies and  smearing effects in the ZEUS detector.
The distributions of generated variables 
were reweighted so as to  reproduce the measured distributions 
after reconstruction. 
Corrections for the data, evaluated on the basis of the Monte Carlo samples,
were calculated independently  in each bin of any given variable.

\begin{figure}
\centerline{
\psfig
{figure=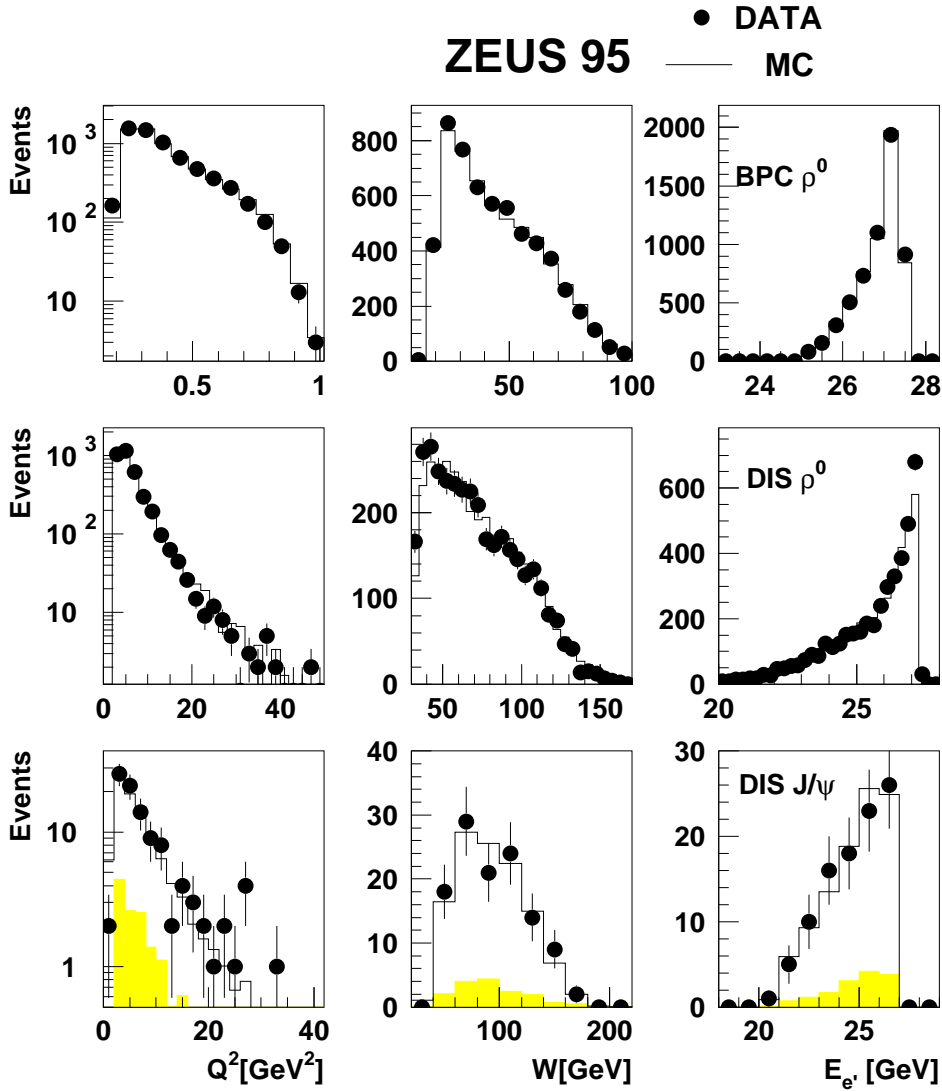,width=\textwidth}
}
\caption{
\label{fig:q2w}
Comparison of the  measured and Monte-Carlo-simulated distributions for 
$Q^2$, $W$  and $E_{{\rm e}^{\prime}}$, the energy of the scattered positron.
The shaded area indicates the  contribution from the Bethe-Heitler process.
}
\end{figure}

\begin{figure}
\centerline{
\psfig
{figure=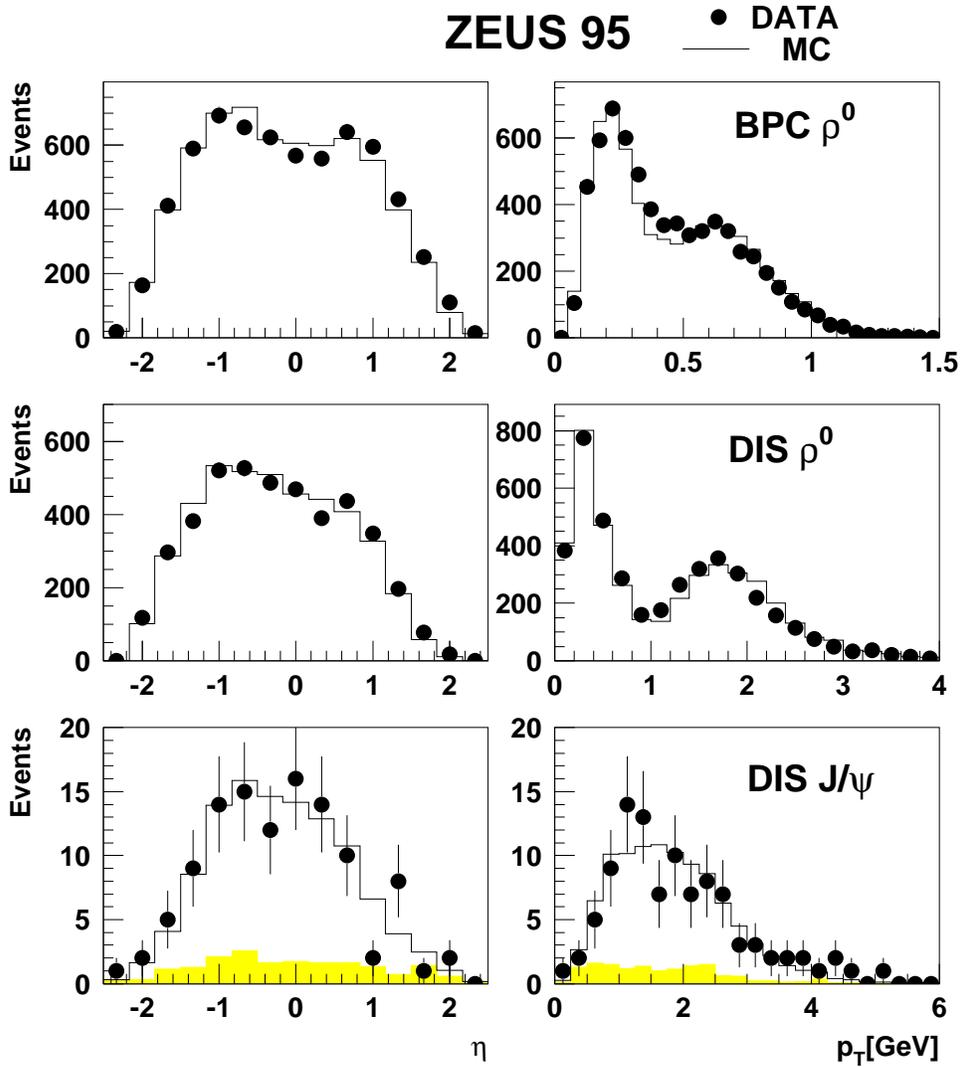,width=\textwidth}
}
\caption{
\label{fig:etapt}
Comparison of the  measured and Monte-Carlo-simulated 
distributions for  pseudorapidity,  $\eta$, and the transverse momentum
in the laboratory frame,  $p_{\rm T}$, 
for the positively charged decay particle. 
The shaded  area indicates the  contribution from the Bethe-Heitler process.}
\end{figure}

\begin{figure}
\centerline{
\psfig
{figure=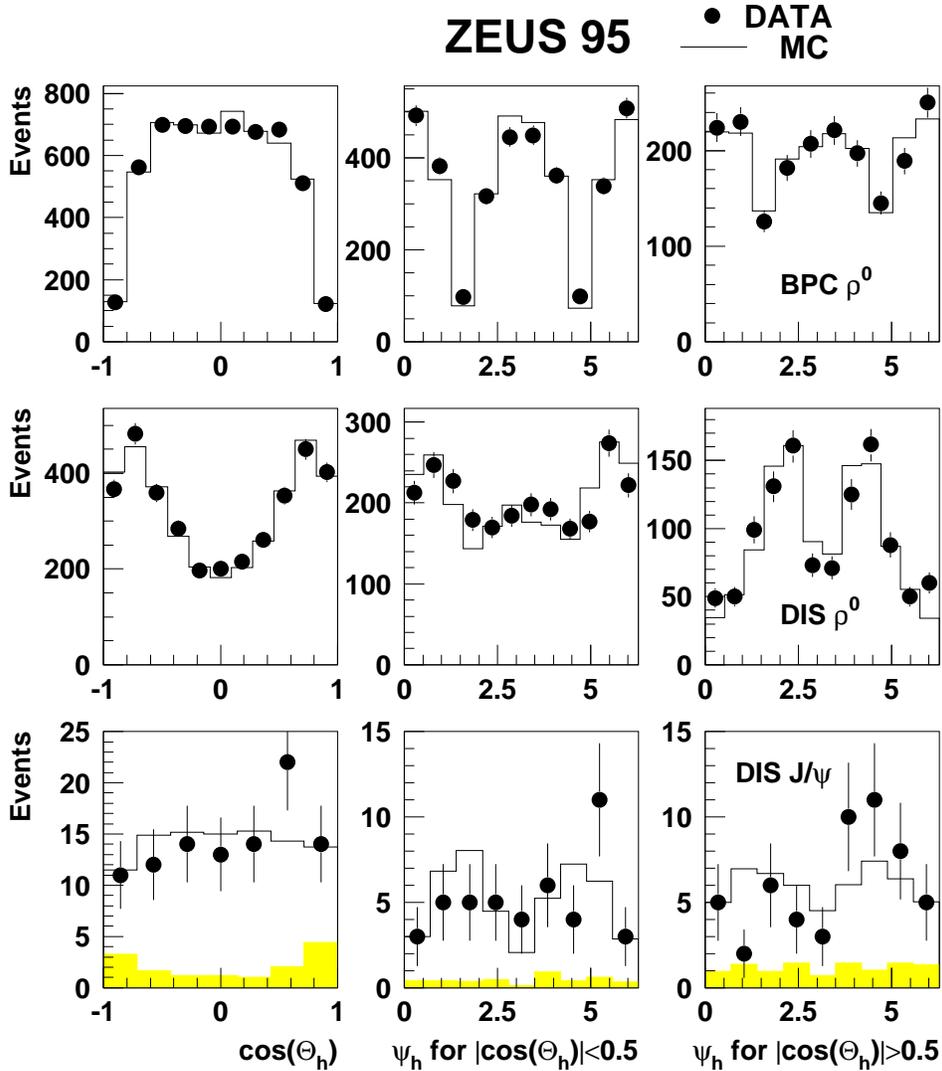,width=\textwidth}
}
\caption{
\label{fig:angles}
Comparison of the  measured and Monte-Carlo-simulated 
distributions of $\cos\theta_{\rm h}$ and $\psi_{\rm h}$. 
The $\psi_{\rm h}$ distributions are shown for two ranges of $\cos\theta_{\rm h}$. 
The strong correlation of these variables is evident.
The shaded area indicates the  contribution from the Bethe-Heitler process.
}
\end{figure}

A comparison of  data and MC simulation 
is presented in Fig.~\ref{fig:q2w} ($Q^2$, $W$ and $E_{{\rm e}^{\prime}}$),
Fig.~\ref{fig:etapt} ($\eta$ and $p_{\rm T}$) and Fig.~\ref{fig:angles} 
($\cos\theta_{\rm h}$ and $\psi_{\rm h}$). The $J/\psi$ sample is restricted to the mass range $2.85 < M_{{\rm l}^+{\rm l}^-}  < 3.25$~{\gev} in order to 
reduce the contribution from hadron pairs.
The dominant remaining background 
originates from  the Bethe-Heitler 
process ${\rm e p} \rightarrow {\rm e}\,{\rm l}^+{\rm l}^-\,{\rm p}$, 
and this contribution is  represented by the shaded areas
in the respective  figures. This process,
where the lepton pairs are either electrons or a muons, proceeds via the fusion of 
a photon radiated by the incoming electron and  one radiated by the proton.
Single particle distributions are very sensitive to  the correct
simulation of the $W$, $Q^2$, $t$ and decay angle  variables. 
As an example,  the transverse momentum distribution 
of the decay   pion  in  the DIS $\rho^0$ sample (Fig.~\ref{fig:etapt})
displays a two-peak structure  with maxima positioned around $\sim 0.3$ and $\sim 1.8$
GeV$^{2}$ (this shape is less distinct in  the case of the BPC $\rho^0$s). 
Since at large values of $Q^2$  the $\rho^0$ mesons are predominantly produced 
in the helicity zero  state,   one of the decay pions is emitted along the direction 
of the $\rho^0$  while the other one is approximately  at rest
in  the $\gamma^{\star} p$ centre-of-mass frame.  This configuration results
in the $p_{\rm T}$ spectrum, measured in the laboratory  frame, shown in Fig.~\ref{fig:etapt}.
The measured and simulated  spectra  of  $\cos\theta_{\rm h}$ and
$\psi_{\rm h}$ (for $|\cos\theta_{\rm h}|<0.5$ and $|\cos\theta_{\rm h}|>0.5$)
are shown in Fig.~\ref{fig:angles}.
The polar and azimuthal  angular distributions are strongly correlated
and the observed agreement between the measured and the simulated
distributions was  obtained via careful tuning of the simulation.


\section{Background}\label{sec:systematics}

After applying the selection criteria described earlier, the data still contain
contributions from various background processes:

\begin{itemize}

\item Proton-dissociative vector-meson production, ${\rm e p}\rightarrow{\rm e V N}$,
where $N$ is a state of mass $M_{\rm N}$ into which the proton diffractively
dissociates. 

\item Elastic production of $\omega$ and $\phi$ mesons (for the $\rho^0$ 
analyses) and of $\psi^{\prime}$ mesons (for the $J/\psi$ analysis).

\item Photon diffractive dissociation, ${\rm e p}\rightarrow{\rm e X p}$ and
${\rm e p}\rightarrow{\rm e X N}$, in which the photon diffractively dissociates 
into  a state $X$ and the proton either remains intact or dissociates.
 
\item Bethe-Heitler production of e$^+$e$^-$ and $\mu^+ \mu^-$ pairs.

\item Beam-gas interactions.

\end{itemize}

The main source of background consists of events with the proton 
diffractively dissociating into hadrons. Some of the hadrons deposit energy 
around the beam pipe in the FCAL but a fraction escape detection. 
The contribution by this process to the observed yields was estimated by 
using exclusive VM events with an energy deposit 
of at least 0.4~{\gev} in the FCAL. (Contamination of this sample by
DIS events is negligible.)

\begin{figure}
\centerline{
\psfig
{figure=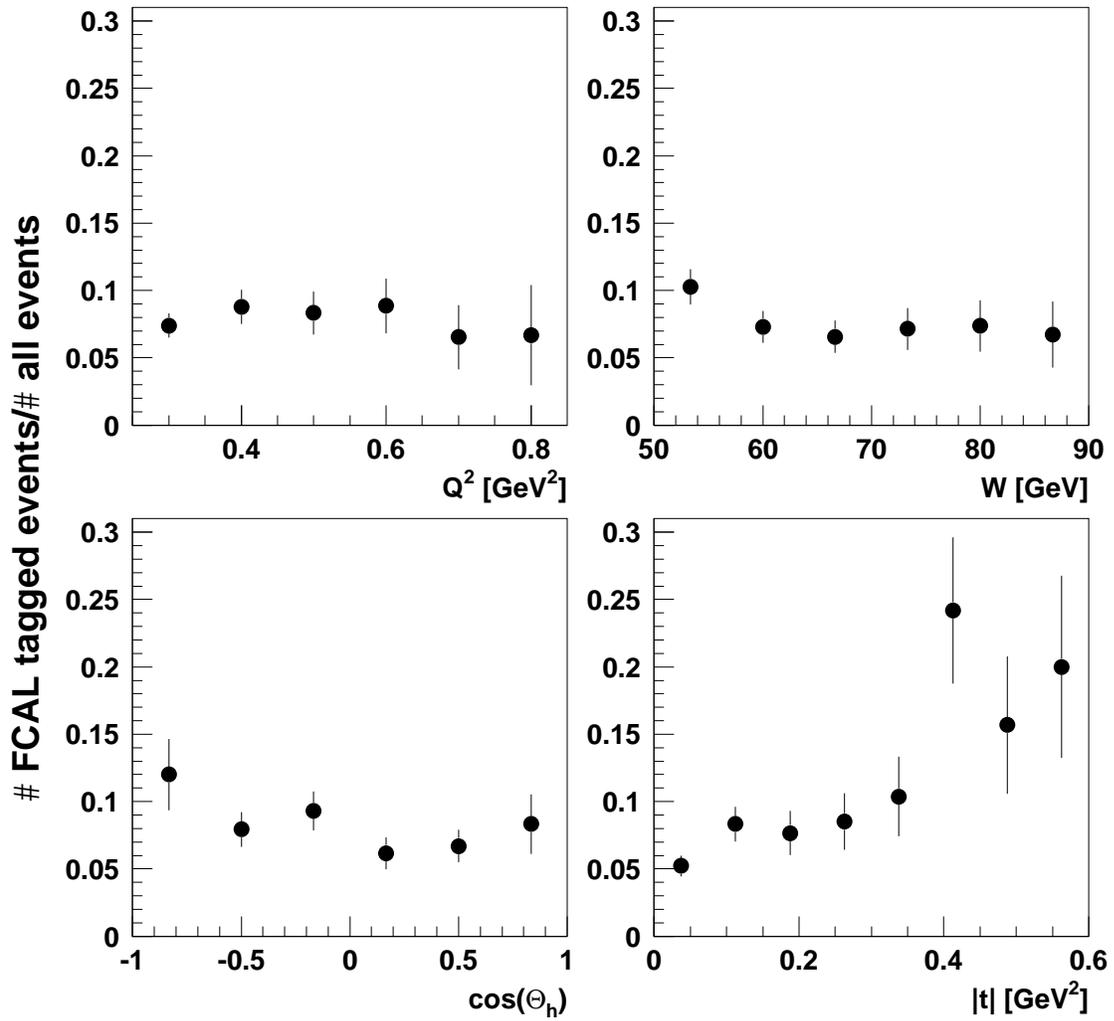,width=\textwidth}
}
\caption{
\label{fig:fcal_bpc}
The fraction  of events tagged with the FCAL as a function of 
$Q^2$, $W$, $\cos\theta_{\rm h}$ and $|t|$ for the BPC $\rho^0$ sample.
}     
\end{figure}

Fig.~\ref{fig:fcal_bpc} shows the ratios of the $Q^2$, $W$, $\cos\theta_{\rm h}$ 
and $t$ distributions for FCAL-tagged events  to those for all events
in the BPC $\rho^0$ sample.  The fraction of  FCAL-tagged events 
is approximately  independent of $Q^2$, $W$ and $\cos{\theta_{\rm h}}$. 
However, a significant dependence on $t$ 
is observed. The latter is expected as a consequence of 
the different $t$ dependences of the cross sections for elastic and 
proton-dissociative reactions. 
The same conclusions can be reached for the  FCAL-tagged events  
from the   DIS  $\rho^0$ and  $J/\psi$ samples.
Under the assumption that a tag in the FCAL  does not affect the 
shape of the acceptance as a function of $Q^2$, $W$ and $\cos{\theta_{\rm h}}$
(as indicated by PYTHIA and EPSOFT),
this result suggests that proton-dissociative and elastic vector-meson production
have the same $Q^2$, $W$ and $\cos{\theta_{\rm h}}$ distributions. This 
supports the hypothesis of factorisation of diffractive 
vertices~\cite{ref:fact}. 
Similar conclusions were reached earlier for $\rho^0$ production
at $Q^2 \simeq 0$~\cite{ref:rhopdissZEUS} 
and in DIS~\cite{ref:rhopdissH1}. 

In the BPC and DIS $\rho^0$ analyses the proton-dissociative background 
evaluation was performed by comparing the number of events with energy 
deposited in FCAL in the data and the proton dissociative EPSOFT or PYTHIA 
samples.
Specifically, the number of residual proton-dissociative events
in the data with FCAL energy smaller than the threshold $E_0$ (1~{\gev} for
the BPC analysis and 0.4~{\gev} for the DIS $\rho^0$ analysis) was estimated as
\begin{eqnarray}
N_{\rm pdiss}^{\rm DATA} = \left\{
\frac{N_{\rm pdiss}} {N(E_{\rm FCAL}>E_0)}
\right\}^{\rm MC}
\times \left\{N(E_{\rm FCAL}>E_0)\right\}^{\rm DATA}, \nonumber
\end{eqnarray}
\noindent
where $N_{\rm pdiss}$ is the fraction of elastic events passing the final cuts.
A total of 160 (64) FCAL-tagged events were used for the BPC $\rho^0$ (DIS $\rho^0$) 
analysis. In the BPC case, the additional 
requirement $W>50$~{\gev} was also imposed, in order to reduce the 
contribution by
nondiffractive events.
The overall contamination integrated up to $|t| = 0.6$~{\gevsq} was
estimated to be $(23 \pm 8 )$\% for the BPC $\rho^0$ 
sample and $(24^{+9}_{-5})$\% for the DIS $\rho^0$ sample,
where the errors represent the total statistical and systematic 
uncertainties.
In the DIS $J/\psi$ analysis, the contamination was found
to be $(21 ^{+10}_{-9})$\%, consistent with the values found in 
the BPC and DIS $\rho^0$ analyses. 

Contamination from elastic production of $\omega$ and $\phi$ mesons 
in the $\rho^0$ analyses was estimated by Monte Carlo studies to be less
than 2\%. Contamination from $\psi^{\prime}$ production in the 
$J/\psi$ analysis is  $(4\pm 1)$\%~\cite{ref:ming}. 
The Bethe-Heitler contribution is approximately 15\%; 
its size was estimated from the LPAIR Monte Carlo simulation~\cite{ref:lpair}.

The photon-dissociative background was studied with a sample of events
generated using PYTHIA. The events which pass the selection criteria
of the present $\rho^0$ analyses have a flat distribution in $M_{\pi\pi}$
up to about $M_{\pi\pi} \simeq 1.4$~{\gev}. If all events at 
$M_{\pi\pi}=1.4$~{\gev} are ascribed to this process, a 3\% upper limit
on the contamination from photon dissociation is deduced. A similar 
result is obtained if an extra constant term is added to the Breit-Wigner
function used for the mass fit.

A contamination of 1.5\% from beam-gas events was deduced from
event samples derived from unpaired electron and
proton bunches, to which all the selection criteria
described earlier were applied.

All subsequent results are shown after subtraction of the contributions from 
proton-dissociative and (for the $J/\psi$ analysis) Bethe-Heitler events.
The estimates of the other backgrounds were included in the systematic 
uncertainties.

\section{Systematic uncertainties} 

The systematic uncertainties are dominated by the uncertainties 
in the acceptance, the proton-dissociative background and the number of 
$\rho^0$ or $J/\psi$ signal events. Table~\ref{syserr} summarises the 
various contributions to the uncertainties in the
integrated cross section for the three analyses. 

In the following sections, whenever a result for a given quantity was obtained 
from a fit, it should be understood that 
the corresponding systematic uncertainty was determined by repeating the fit 
for each systematic check. The differences between the values of the 
quantity thus found and its nominal value were added in quadrature.

The trigger efficiency and its uncertainty were estimated, whenever possible, 
by using samples of events selected by independent triggers.
The model dependence was investigated by comparing the acceptances obtained with 
various Monte Carlo generators, or with the same generator
but with different input parameters. In particular, the sensitivity to the $W$ 
and $Q^2$ dependences of the cross section in the generator were studied,
as well as the sensitivity to $R$.
Various electron finders were used to estimate the uncertainty due to 
the electron identification in the DIS analyses. In the BPC case 
a significant 
contribution originates from the uncertainty in the BPC alignment; its effect 
increases with decreasing $Q^2$. The sensitivity to the cuts
mostly reflects the effect of the CTD-CAL-matching  and 
track quality requirements.

The contribution due to the extraction of the number of  signal  events, reflecting
the sensitivity to the mass fitting procedure, has been discussed  
previously~\cite{ref:rhoZEUS93}  for the $\rho^0$ analyses. 
In the $J/\psi$ analysis, this uncertainty 
is dominated by the sensitivity to the shape used for the subtraction
of the nonresonant background.

The uncertainties in the luminosity, trigger efficiency, photon
flux determination, $\omega$, $\phi$ and $\psi^{\prime}$ backgrounds,
photon dissociation (for $\rho^0$s), proton dissociation, 
beam-gas contamination, and $J/\psi$
decay branching ratios  are treated as overall normalisation uncertainties.
In addition, for the BPC {$\rho^0$} and $J/\psi$ analyses, the contributions
from the signal extraction procedure and from radiative corrections 
were included
in the normalisation uncertainty. The normalisation uncertainties are 
$^{+9\%}_{-14\%}$ for DIS $\rho^0$, $^{+15\%}_{-16\%}$ for BPC $\rho^0$, and
$^{+13\%}_{-15\%}$ for the $J/\psi$, dominated by proton dissociation and, for
the BPC $\rho^0$, the signal extraction.


\section{Results}

\subsection{Mass distributions}

\begin{figure}
\centerline{
\psfig
{figure=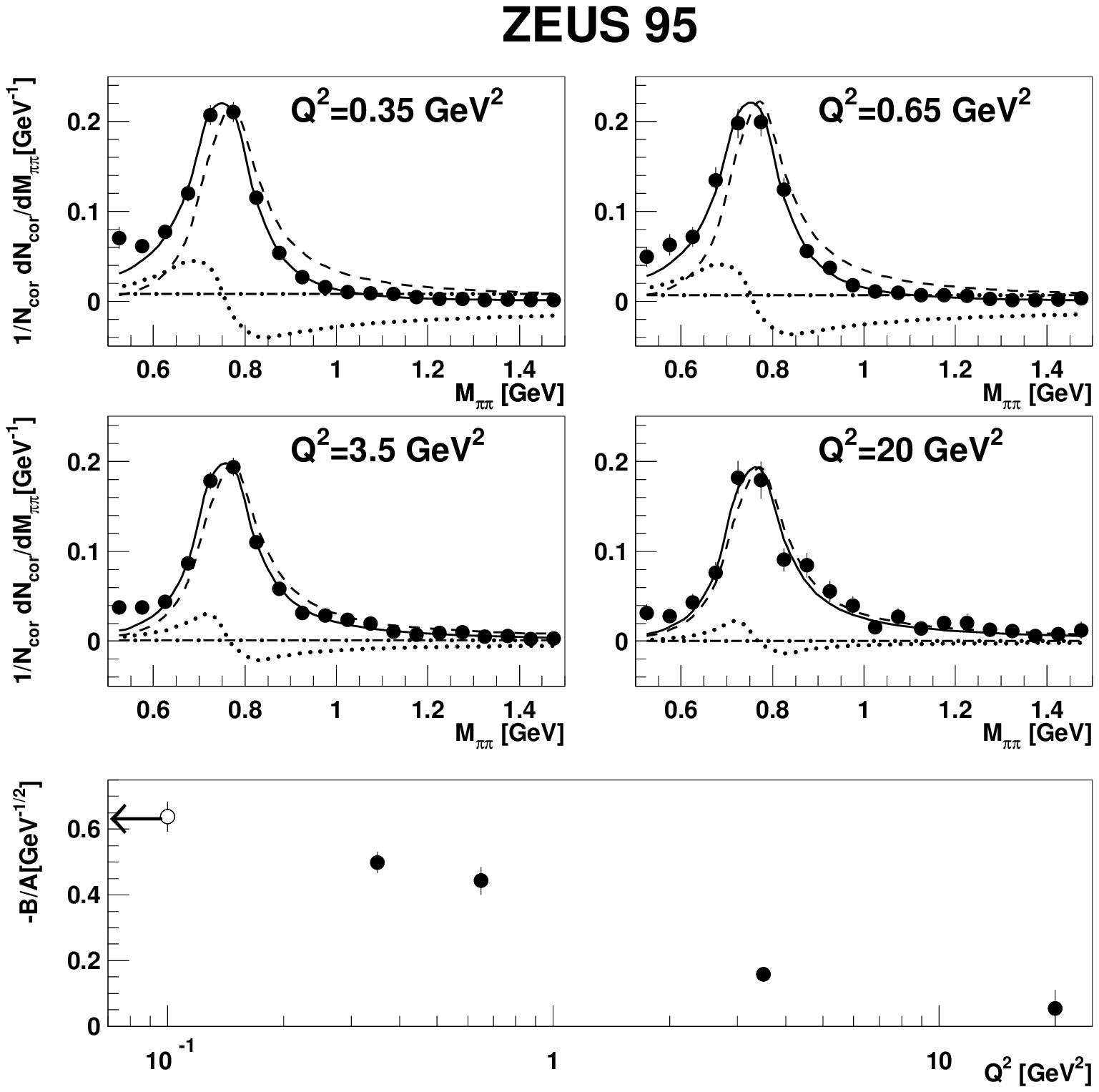,width=\textwidth}
}
\caption{
\label{fig:mass_rho}    
Top: The acceptance-corrected
differential mass distributions, normalised to unit area,
for the BPC and DIS $\rho^0$ samples. Line types:
solid  -- fit based on the S\"{o}ding 
model~\protect{\cite{ref:soeding}}(cf. Eq.~\protect{\ref{eq:soeding}});
dashed -- contribution from the p-wave Breit-Wigner term;
dotted  -- interference term;
dash-dotted  -- background contribution.
Bottom: ratio $B/A$ (cf. Eq.~\protect{\ref{eq:soeding}}).
The open point associated with an arrow indicates 
the value measured in photoproduction~\protect{\cite{ref:rhopdissZEUS}}.
}
\end{figure}

Acceptance-corrected  differential distributions ${\rm d}N/{\rm d}M_{\pi^+\pi^-}$ 
for  BPC and DIS $\rho^0$ samples  are shown in Fig.~\ref{fig:mass_rho}.
The $\pi^+\pi^-$ mass spectra deviate  from the shape of 
a relativistic p-wave Breit-Wigner function.  
This effect may be explained by the  interference
between nonresonant and resonant $\pi^+\pi^-$
production amplitudes~\cite{ref:soeding}.  
The differential distributions ${\rm d}N/{\rm d}M_{\pi^+\pi^-}$ were fitted in the 
range  $0.6<M_{\pi^+\pi^-}<1.2$~{\gev}, in several $Q^2$ intervals,  
using a parameterisation based on the 
S\"{o}ding model~\cite{ref:soeding},  which  accounts for the effect of interference
according to~\cite{ref:rhopdissZEUS}
\begin{equation}\label{eq:soeding}           
{\rm d}N/{\rm d}M_{\pi\pi} = \Big| A\frac{\sqrt{M_{\pi\pi}M_{\rho}\Gamma_{\rho}}}
{M^2_{\pi\pi}-M^2_{\rho}+iM_{\rho}\Gamma_{\rho}} + B\Big|^{2},
\end{equation}
where $M_{\rho}$ and $\Gamma_{\rho}$ are the nominal mass and  width
of the $\rho^0$ meson, respectively; 
$B$ is the nonresonant amplitude  assumed to be constant and real and
$A$ is a  normalisation constant. 
The values of the $\rho^0$ meson mass and width obtained
by fitting Eq.~\ref{eq:soeding} are 
$768\pm3$(stat.)~{\mev} and $152\pm6$(stat.)~{\mev} for the  BPC $\rho^0$  sample  
and $762\pm3$(stat.)~{\mev} and $146\pm7$(stat.)~{\mev}  for the DIS 
$\rho^0$  sample.
The ratio $B/A$  decreases with $Q^2$, as  shown 
in Fig.~\ref{fig:mass_rho}.
\par
\begin{figure}
\centerline{
\psfig
{figure=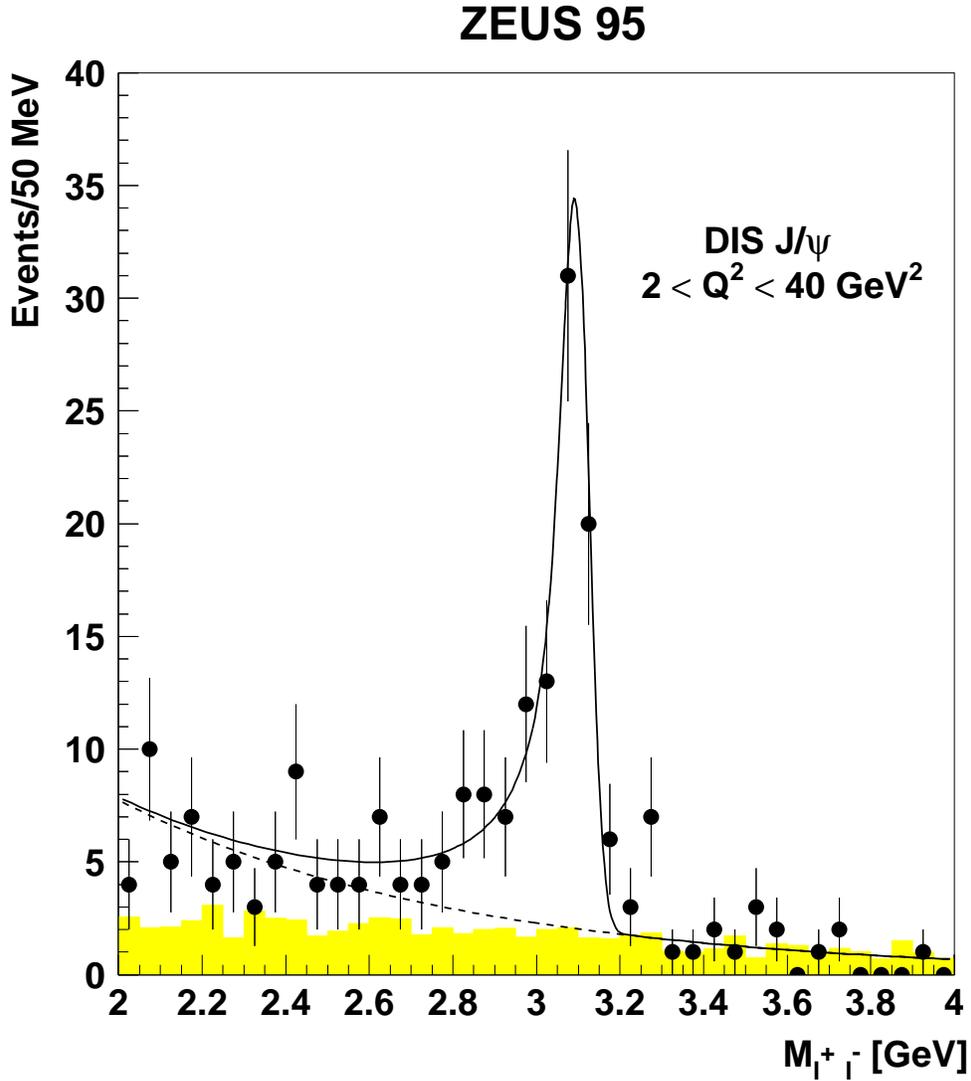,width=\textwidth}
}
\caption{
\label{fig:mass_jpsi}
The differential mass distribution 
for the  DIS $J/\psi$ sample (not corrected for acceptance).
The distribution includes contributions from both $e^+e^-$ and
$\mu^+\mu^-$ pairs. Solid line -- signal function (see text), 
dashed line -- background; shaded  area -- 
contribution from the Bethe-Heitler process.
}     
\end{figure}

\medskip\noindent
The uncorrected differential distribution  ${\rm d}N/{\rm d}M_{{\rm l}^+{\rm l}^-}$  
for the  DIS $J/\psi$ 
sample,  shown in Fig.~\ref{fig:mass_jpsi},
was   fitted  in the  range  $2 < M_{{\rm l}^+{\rm l}^-} < 4$~{\gev}
with the sum of a  signal function  and an  exponentially falling  background.
The former is   a convolution of a Gaussian resolution function
with the  $J/\psi$ mass spectrum  obtained using  the DIPSI Monte Carlo 
generator~\cite{ref:dipsi} including bremsstrahlung. No positive muon or electron identification was performed; the muon mass was used in calculating 
$M_{{\rm l}^+{\rm l}^-}$.
The main contributions to the background are from  oppositely 
charged hadrons and from the Bethe-Heitler process.
The fitted  value of the $J/\psi$ 
mass is   $3.114\pm 0.006$(stat.)~{\gev} and the  width of the Gaussian 
resolution function is 26~{\mev}. 
Integrating the  fitted function in the above  $M_{{\rm l}^+{\rm l}^-}$  range
yields  a signal of  $97\pm12$ $J/\psi$ mesons.

\subsection{Total  cross sections}

The total cross sections for exclusive $\rho^0$ and $J/\psi$
electroproduction, ${\rm e p}\rightarrow{\rm e V p}$,
were  determined  using  the expression
\begin{equation}\label{eq:sigma_tot}
\sigma({\rm e p} \rightarrow {\rm e V p}) = \frac{N \cdot \Delta}
                                     {A \cdot {\cal L}},
\end{equation}
where $N$ is the number of events in data,  
$A$  the  overall acceptance, ${\cal L}$  the integrated luminosity and 
$\Delta$ the correction for the  proton dissociation background.
For the $\rho^0$ we quote
the integrated cross sections for $|t|<0.6~{\gevsq}$ and 
for the invariant  mass range 
$2m_{\pi} < M_{\pi\pi} < M_{\rho} + 5 \Gamma_{\rho}$,
where $m_{\pi}$ is the mass of a charged pion, $M_{\pi\pi}$ is the invariant mass
of the two pions, $M_{\rho}$ is the nominal $\rho^0$ mass and $\Gamma_{\rho}$ 
is the width of the $\rho^0$ resonance at the nominal $\rho^0$ mass.
In the DIS $\rho^0$ analysis, we correct to the Born level.

Total cross sections for exclusive $\rho^0$ and $J/\psi$
production, ${\rm ep} \rightarrow {\rm eV p}$  and  
$\gamma^{\star} {\rm p} \rightarrow {\rm V p}$, 
are given  in Tables~\protect{\ref{tab:xs_rho}} and~\protect{\ref{tab:xs_jpsi}}.
The cross sections in each $Q^2$ and $W$ interval are quoted  at values
close to the weighted  averages in the bins. 
The $\gamma^{\star} {\rm p}$ cross sections  were obtained from the ep
cross sections using formulae~(\ref{eq:d2epdydq2})-(\ref{eq:sigmatot}).
They are insensitive to the value of $R$ since $\epsilon\simeq$1.

\subsection{$Q^2$ dependence} \label{sec:q2dep}

\begin{figure}
\centerline{
\psfig
{figure=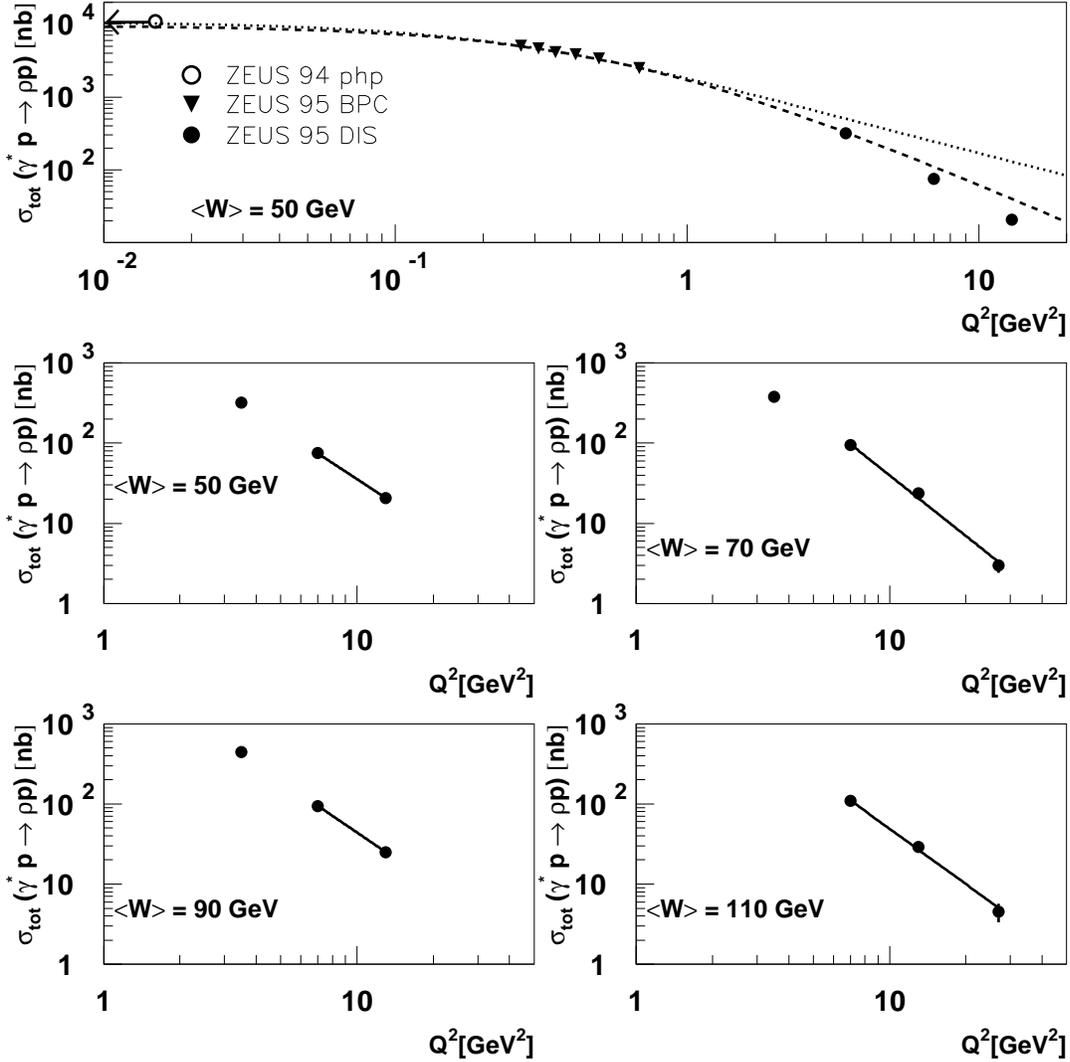,width=\textwidth}
}
\caption{
\label{fig:q2dep_rho}
The cross sections, $\sigma_{\rm tot}^{\gamma^{\star}{\rm p}}$,
for exclusive $\rho^0$ production 
as a function of $Q^2$ for various values of $W$. Top figure: 
the curves  represent fits to the low-$Q^2$ (BPC) data using the functions
$(1+R)/(1+Q^2/M_{\rm eff}^2)^2$ (dotted line) and 
$1/(1+Q^2/M_{\rho}^2)^n$ (dashed line).
The open point with the horizontal arrow indicates 
the value measured in photoproduction~\protect{\cite{ref:rhopdissZEUS}}.
Four bottom figures: the solid lines represent a fit 
of the form $\sigma_{\rm tot}^{\gamma^{\star}{\rm p}} \propto Q^{-2n}$
for $Q^2>5$~{\gevsq}.
Only statistical uncertainties are shown.
}
\end{figure}

Fig.~\ref{fig:q2dep_rho}  shows the cross section for the process 
$\gamma^{\star}{\rm p} \rightarrow \rho^0 {\rm p}$ as a function of $Q^2$.
The  low-$Q^2$ data from this analysis have been fitted with two  
VMD-motivated functions and the corresponding  curves are shown 
in the upper plot. A fit to the function 
$\sigma_{\rm tot}^{\gamma^{\star}{\rm p}} \propto [1+R(Q^2)]/(1+Q^2/M_{\rm eff}^2)^2$,
yielded   $M_{\rm eff}=0.66\pm0.05$(stat.)$\pm0.10$(syst.)~{\gev}
(here  $R=\sigma_{\rm L}^{\gamma^{\star}{\rm p}} / \sigma_{\rm T}^{\gamma^{\star}{\rm p}}$ was taken as measured in  this analysis). 
A fit to   $\sigma_{\rm tot}^{\gamma^{\star}{\rm p}} \propto 1/(1+Q^2/M_{\rho}^2)^n$ gives
$n=1.75\pm$0.10(stat.)$\pm$0.29(syst.) for the entire sample ($W_0$=50~{\gev}).

The $\gamma^{\star} {\rm p}$ cross sections are shown for the DIS data for fixed $W$ as a 
function of $Q^2$ in the four lower plots.  The cross section measurement
at $Q^2=27(13)$~{\gevsq} and $W=80(120)$~{\gev} has been translated
to $W=70(110)$~{\gev} using  the $W$ dependence measured in this analysis.  
The data are consistent with a  simple power law behaviour for $Q^2>5$~{\gevsq}.  
Fitting the points at $Q^2>5$~{\gevsq} with the form $Q^{-2n}$ yields 
$n=2.07\pm0.22, \; 2.51\pm0.15, \; 2.15\pm 0.31, \; 2.29\pm 0.18$ 
for $W=50,70,90,110$~{\gev}, where the  errors are  statistical  only.
The systematic uncertainties are typically 0.05.
The $Q^2$ dependence is consistent with being  independent of $W$,
and averaging the four values yields $n=2.32\pm0.10$(stat.).

\begin{figure}
\centerline{
\psfig
{figure=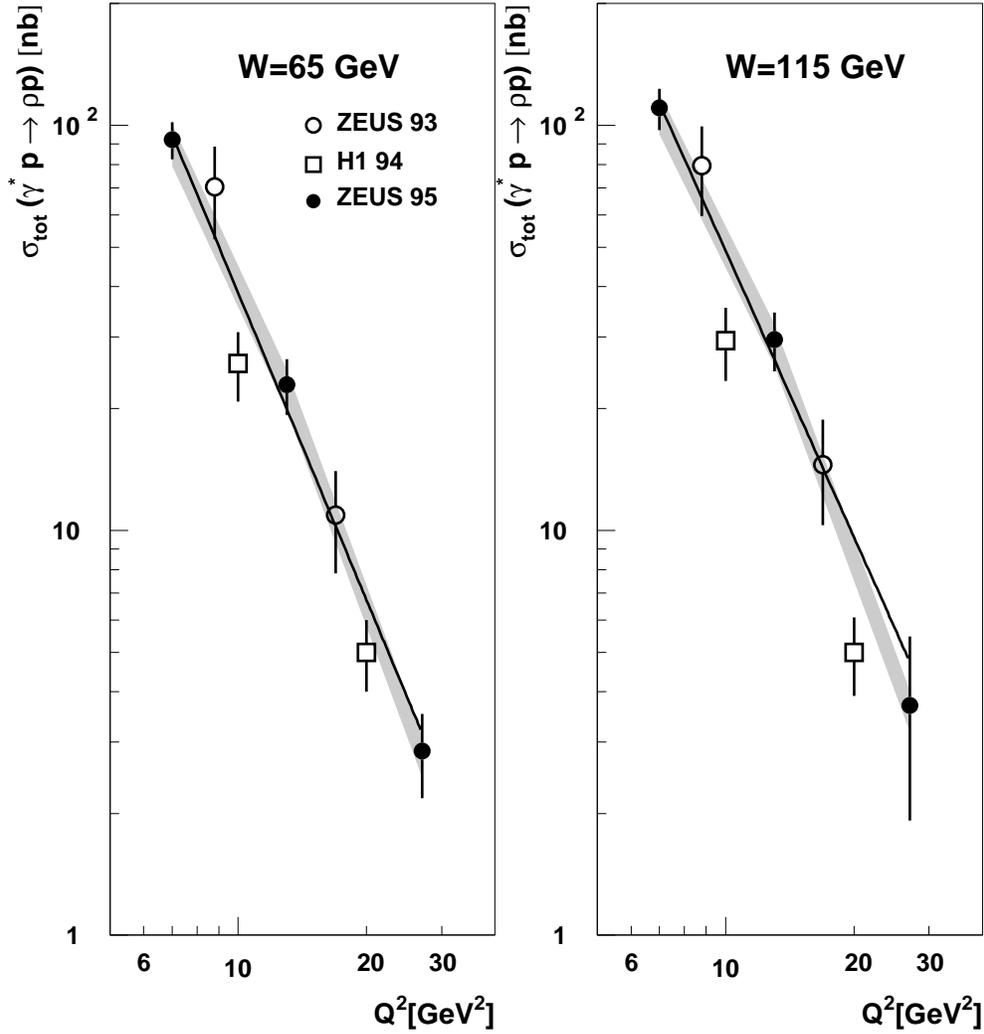,width=\textwidth}
}
\caption{
\label{fig:q2comp}
Comparison of the $Q^2$ dependence of $\sigma_{\rm tot}^{\gamma^{\star}{\rm p}}$
measured by ZEUS and H1. The ZEUS points were moved to the $W$ values quoted
for the H1 measurements. The lines represent the results of fits to ZEUS 95 data.
The error
bars on the ZEUS points represent the quadratic sum of
statistical and systematic  uncertainties. The normalisation uncertainties in
the ZEUS measurements are indicated by the shaded areas.
}
\end{figure}

The data from this analysis are compared to previous HERA measurements 
in Fig.~\ref{fig:q2comp}.  The cross sections are quoted at the $W$ values 
used by the H1 collaboration~\cite{ref:h1dis}.  
A comparison at fixed $W$ entails smaller translation uncertainties than 
a comparison at fixed $Q^2$, as the $W$ dependence is much weaker than
the $Q^2$ dependence.  A fit to the 95 ZEUS DIS data is shown to guide 
the eye. The results from this analysis are 
in excellent  agreement with the previous ZEUS results~\cite{ref:disvm2}.  
The H1 data  are systematically lower than  the  ZEUS 
measurements by approximately 30 to 40\%.  

\begin{figure}
\centerline{
\psfig
{figure=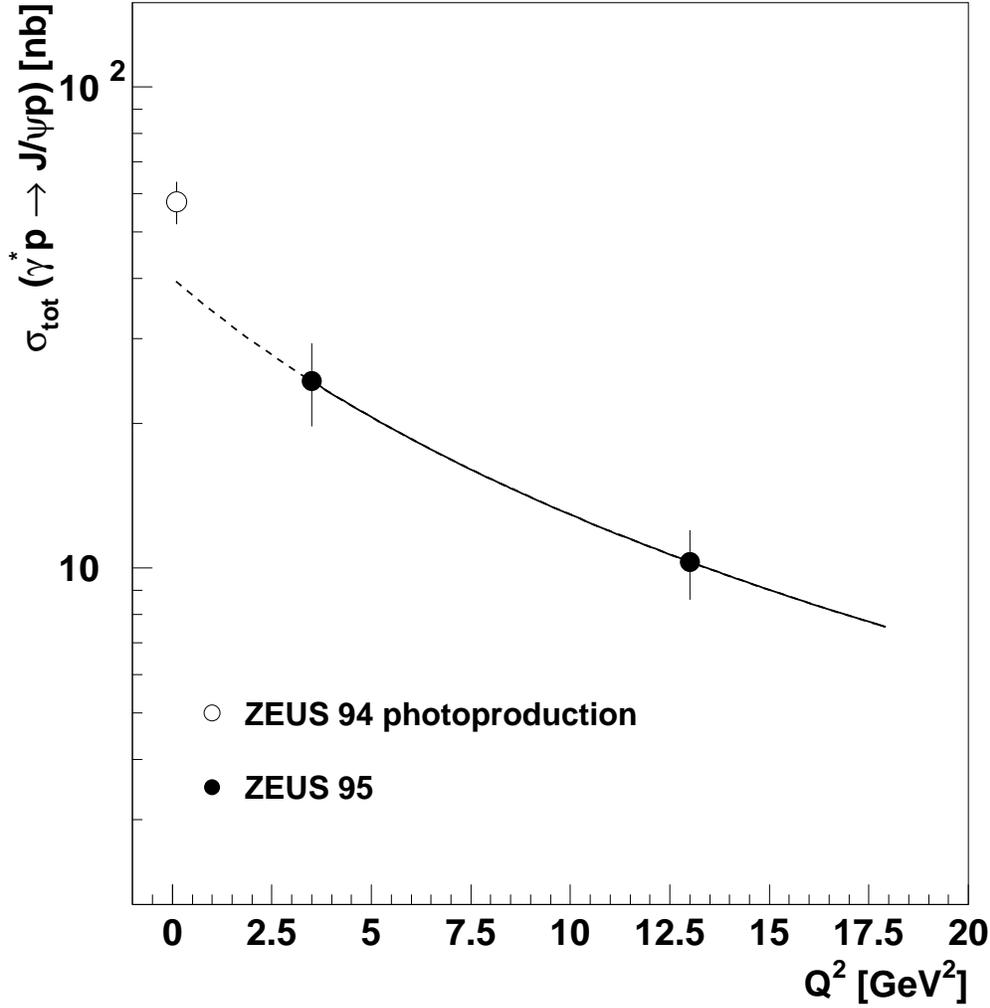,width=\textwidth}
}
\caption{
\label{fig:q2dep_jpsi}
The cross section, $\sigma_{\rm tot}^{\gamma^{\star}{\rm p}}$, 
for exclusive  $J/\psi$ production as a function of $Q^2$.
Line types: solid --  function
$\sigma_{\rm tot}^{\gamma^{\star}{\rm p}} \propto 1/(1+Q^2/M_{J/\psi}^2)^n$ fitted to the  
data points at $Q^2$=3 and 13~{\gevsq};
dashed --   extrapolation of the fit result
to $Q^2$=0. Only statistical errors are shown.
}
\end{figure}

A commonly adopted form for the $Q^2$ dependence of the $J/\psi$
cross section is $\sigma_{\rm tot}^{\gamma^{\star}{\rm p}}  \propto 1/(1+Q^2/M_{J/\psi}^2)^n$. This form
was fitted to the  $\gamma^{\star}{\rm p} \rightarrow J/\psi\; {\rm p}$ cross section 
shown in Fig.~\ref{fig:q2dep_jpsi}. The curve  on the figure 
represents this function fitted to the two measured data points  
(evaluated at $W_0$=90~{\gev}) 
yielding  $n$=1.58$\pm$0.22(stat.)$\pm$0.09(syst.).
This result is consistent with the H1 measurement of
$n=1.9\pm$0.3(stat.)~\cite{ref:h1dis}.

\subsection{$W$ dependence}
\begin{figure}
\centerline{
\psfig
{figure=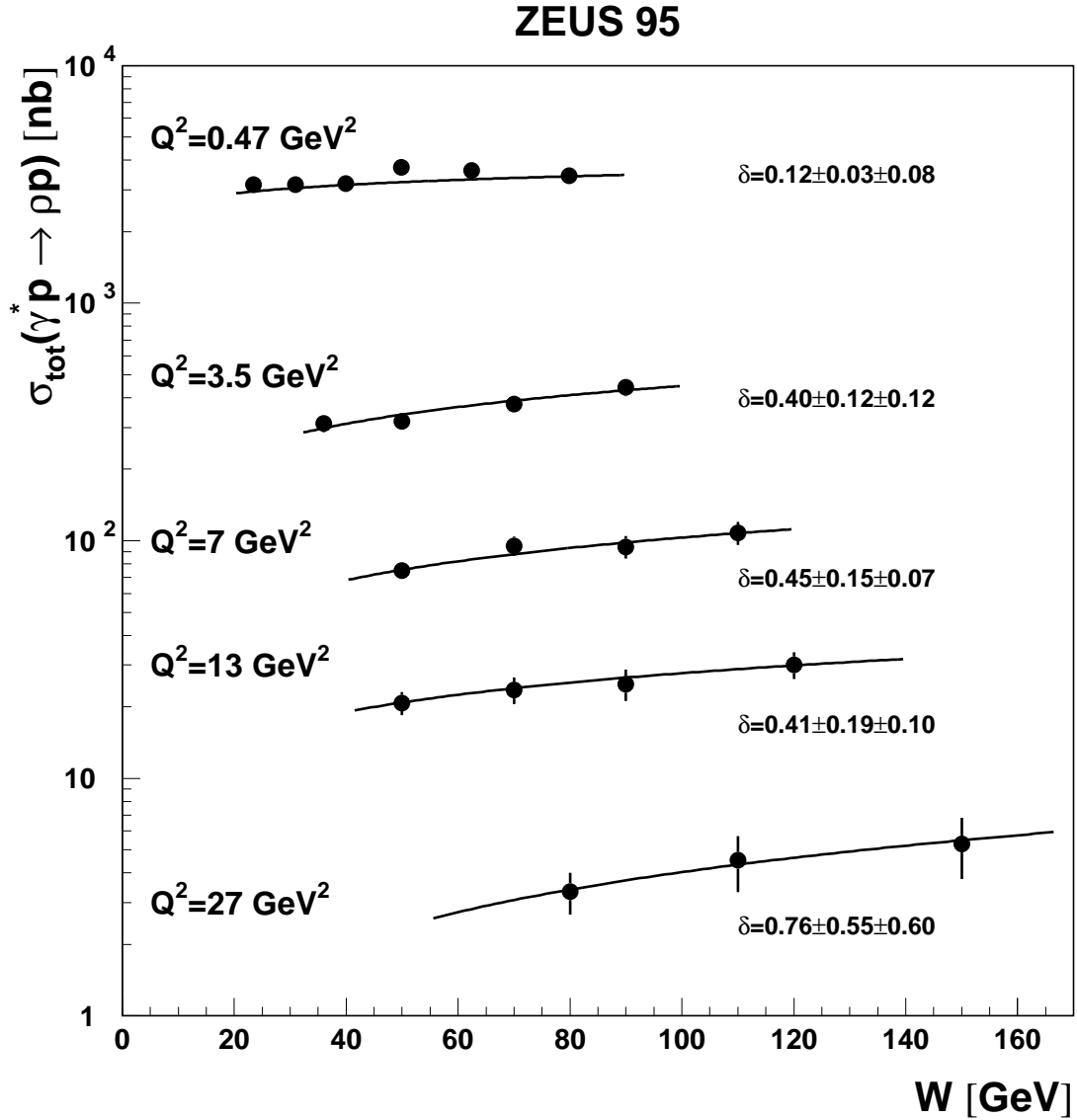,width=\textwidth}
}
\caption{
\label{fig:q2wdep_rho}
The cross section, $\sigma_{\rm tot}^{\gamma^{\star}{\rm p}}$, for exclusive  $\rho^0$ production  
as a function of $W$ for various values of $Q^2$.
Only statistical errors are shown. 
The lines  represent  the fitted parameterisation
$\sigma_{\rm tot}^{\gamma^{\star}{\rm p}} \propto  W^{\delta}$.
}

\end{figure}

\begin{figure}
\centerline{
\psfig
{figure=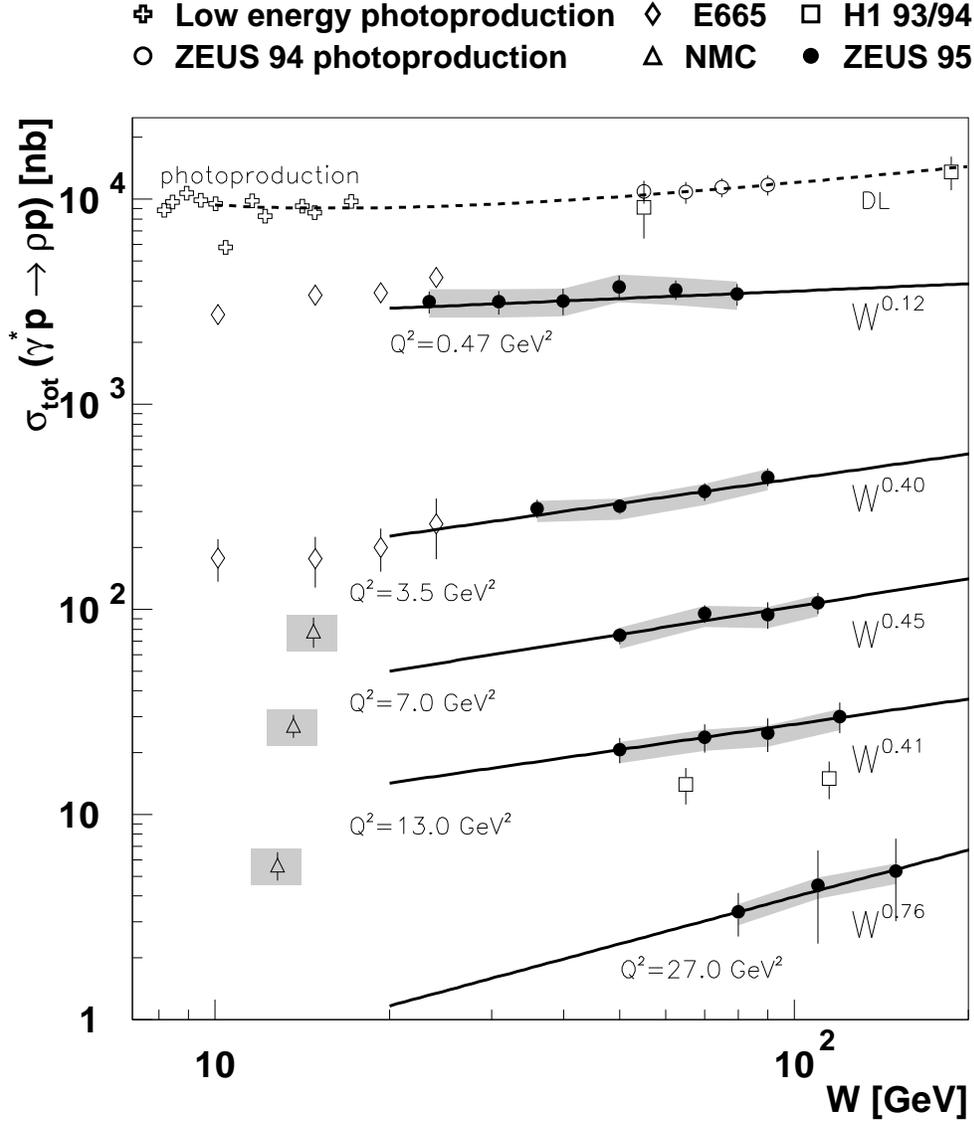,width=\textwidth}
}
\caption{
\label{fig:q2wdep_rho_comp}
Comparison of cross sections, $\sigma_{\rm tot}^{\gamma^{\star}{\rm p}}$, for  exclusive $\rho^0$ production,
as a function of $W$ for various values of $Q^2$. 
The error bars represent statistical and systematic errors added in quadrature.
The solid lines represent the fit results shown 
in Fig.~\ref{fig:q2wdep_rho}.
The dashed line is the prediction by Donnachie 
and Landshoff~\protect{\cite{ref:dl}}.
The overall normalisation uncertainties are shown as shaded bands for the
NMC and ZEUS data points.
The NMC~\protect{\cite{ref:nmc}},  E665~\protect{\cite{ref:e665}}
and H1~\protect{\cite{ref:h1dis}}  data points  
were interpolated  to the indicated  $Q^2$ values (see text).
}
\end{figure}

The  measured cross section for exclusive $\rho^0$ production
as a function of $W$ for  $Q^2$=0.47, 3.5, 7, 13 and 27~{\gevsq} 
is presented in Fig.~\ref{fig:q2wdep_rho}. 
The curves  show the results  of fits to the data using the
function  $ W^{\delta}$. The results of the fits  are given
in Table~\ref{tab:delta}.
The $W$ dependence of the $\rho^0$ production cross section
at low values of $Q^2$ (BPC $\rho^0$) rises slowly with $W$:
$\delta$=0.12$\pm$0.03(stat.)$\pm$0.08(syst.)
for $Q^2$=0.47~{\gevsq}.
This  result  is consistent with the value 
$\delta=0.16\pm0.06$(stat.)$^{+0.11}_{-0.15}$(syst.)
measured 
in  photoproduction~\cite{ref:rhopdissZEUS}. 
Averaging the data in the range $3<Q^2<20$~{\gevsq}
yields  $\delta=0.42\pm 0.12$(stat.$\oplus$syst.), which indicates  
that  the $W$ dependence   increases   with  $Q^2$.
In Fig.~\ref{fig:q2wdep_rho_comp}, the ZEUS data are compared to results
from the NMC~\cite{ref:nmc}, E665~\cite{ref:e665}, and H1~\cite{ref:h1dis} 
experiments.
The NMC, E665 and H1 data points  have been moved  
to coincide with the  $Q^2_0$ values  of   the present  analysis.
This was done according to the $Q^2$  dependence reported by each of the experiments.
The  H1 points at $Q^2$=13~{\gevsq} were obtained by translating
the cross sections measured at  $Q^2$=10~{\gevsq}  and $Q^2$=20~{\gevsq} and taking
a weighted average.
The NMC measurements were moved from $Q^2$=6.9 to  $Q^2$=7.0~{\gevsq} and 
from $Q^2$=11.9 to  $Q^2$=13~{\gevsq}, using values of 
$R$ from the model of Martin, Ryskin and Teubner~\cite{ref:mrt}
to  evaluate  $\sigma_{\rm T}^{\gamma^{\star}{\rm p}} + \sigma_{\rm L}^{\gamma^{\star}{\rm p}}$. The
 E665 measurements were moved from $Q^2$=0.61 to  $Q^2$=0.47~{\gevsq} and 
from $Q^2$=5.69 to $Q^2$=3.5~{\gevsq}.

\begin{figure}
\centerline{
\psfig
{figure=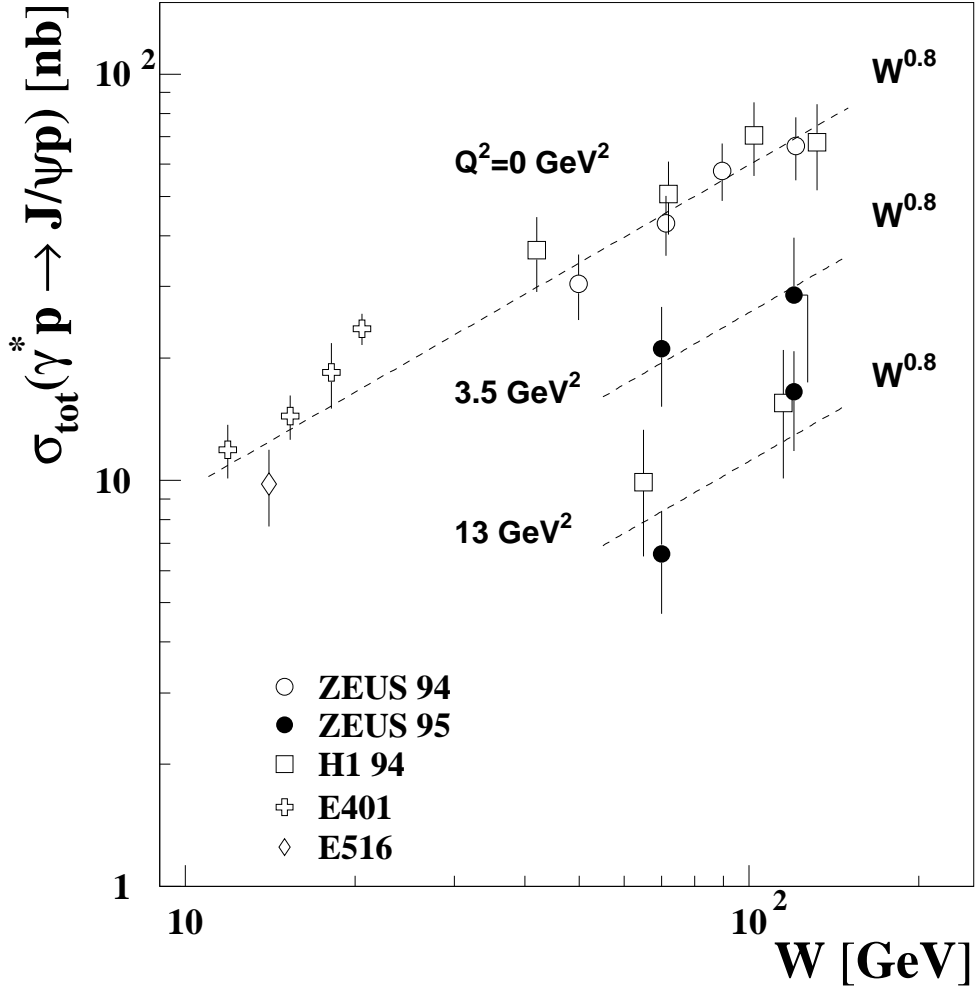,width=\textwidth}
}
\caption{
\label{fig:q2wdep_jpsi_comp}
The cross section, $\sigma_{\rm tot}^{\gamma^{\star}{\rm p}}$, 
for exclusive  $J/\psi$ 
production as a function of $W$ for various values of $Q^2$.
The error bars represent statistical and systematic  uncertainties  
added in quadrature, including the normalisation uncertainty in the ZEUS
measurements of $^{+13\%}_{-15\%}$.
Measurements from the fixed target experiments,
E401~\protect{\cite{ref:e401}} and E516~\protect{\cite{ref:e516}} are 
included for comparison.
The lines, drawn to guide the eye, correspond to
the cross section  parameterisation 
$\sigma_{\rm tot}^{\gamma^{\star}{\rm p}} \propto W^{0.8}$.
}
\end{figure}

The  cross section for  exclusive $J/\psi$ photoproduction,
measured at HERA and at low energies~\cite{ref:e401,ref:e516}, 
shows a rapid rise  with $W$, approximately as $W^{0.8}$.
The measured $W$ dependence  at   higher  values of $Q^2$  is also consistent 
with  this behaviour,  as can be seen  in Fig.~\ref{fig:q2wdep_jpsi_comp}.
In  this figure, the H1 data points were scaled from $Q^2$=16 to 13~{\gevsq}.
The curves, drawn to guide the eye, display a $W^{0.8}$ dependence.

\subsection{Ratio of $J/\psi$ and $\rho^0$ cross sections}

\begin{figure}
\centerline{
\psfig
{figure=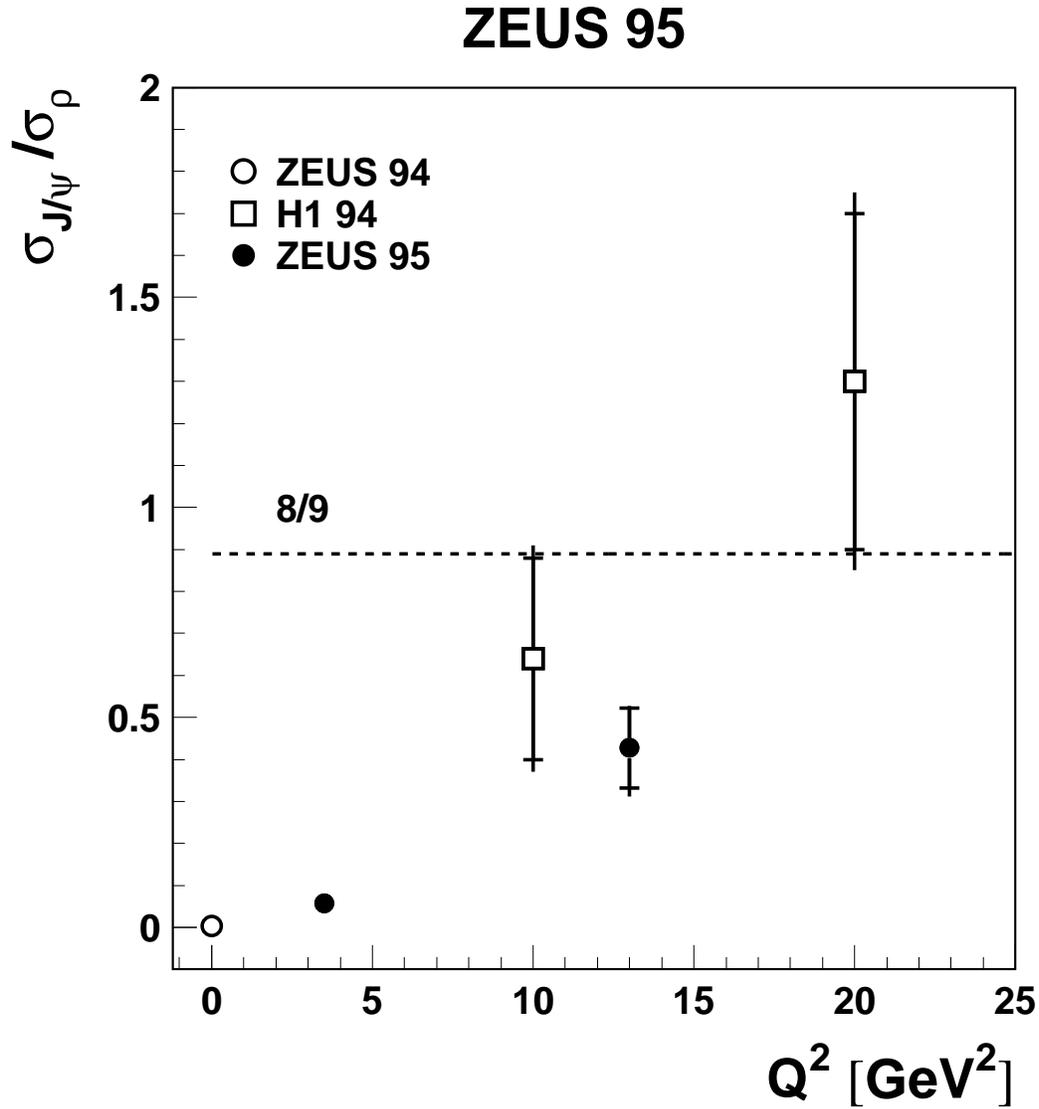,width=\textwidth}
}
\caption{
\label{fig:RATIOjpsirho}
The ratio of the $\gamma^{\star} {\rm p}$ cross sections  for exclusive 
$J/\psi $ 
and $\rho^0$ production as a function of $Q^2$ \cite{ref:zeuselpsi, ref:h1dis}.
The dashed line indicates the flavour-symmetric expectation of 8/9. 
The inner error
bars represent statistical
uncertainties; the outer error bars indicate the quadratic sum of
statistical and systematic  uncertainties.
}
\end{figure}

The values of  the ratio of $J/\psi$ and $\rho^0$ ($\gamma^{\star} {\rm p}$) cross sections, measured at $Q^2$=3.5 and 13~{\gevsq},  are  given  in 
Table~\ref{Tab:RATIOjpsirho}. The correlated errors,  which include those 
associated with the proton dissociation  background subtraction, 
with the uncertainties in the trigger
efficiency and with the uncertainty in the determination of the luminosity,
do not contribute to the uncertainty in the ratio.
The ratio increases with $Q^2$,  as can be seen 
in Fig.~\ref{fig:RATIOjpsirho}.

\subsection{Differential  cross sections ${\rm d}\sigma^{\rm ep}/{\rm d}|t|$}

\begin{figure}
\centerline{
\psfig
{figure=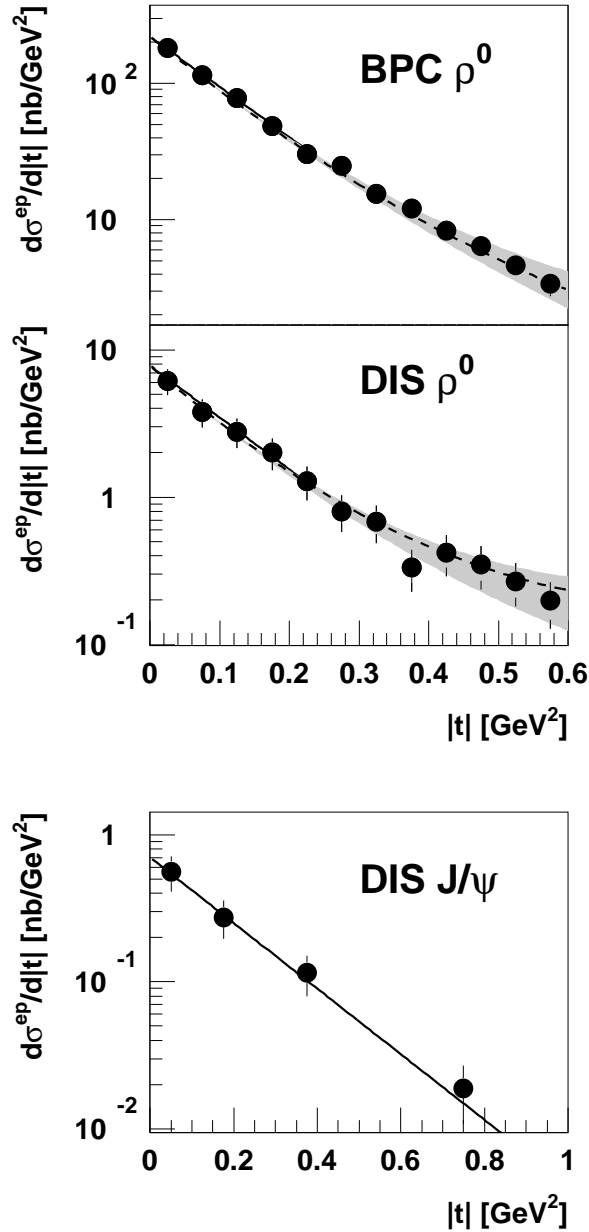,height=20.0cm,width=10.0cm}
}
\caption{
\label{fig:dNdt}
The differential cross section ${\rm d}\sigma^{\rm ep}/{\rm d}|t|$ for
$\pi^+\pi^-$ \mbox{($0.6<M_{\pi\pi}<1.2$~{\gev})}  and $J/\psi$ production. 
Only statistical errors are shown.
Line types: solid -- fit of the function ${\rm d}\sigma^{\rm ep}/{\rm d}|t| \propto e^{-b|t|}$ 
for $|t|<0.3$~{\gevsq} ($|t|<1$~{\gevsq} for $J/\psi$); 
dashed  --  ${\rm d}\sigma^{{\rm ep}}/{\rm d}|t| \propto e^{-b|t|+ct^2}$  for $|t|<0.6$
GeV$^2$; the shaded band  indicates the systematic error resulting
from the uncertainty in the parameter $c$. 
}
\end{figure}

The differential  cross sections
for  exclusive $\rho^0$ production,  ${\rm d}\sigma^{\rm ep}/{\rm d}|t|$,
were measured in several $Q^2$  and  $W$  intervals;
they are shown in Fig.~\ref{fig:dNdt} for the full data samples.
The distributions   were fitted with an  exponential  
function of the form  exp$(-b|t|)$ for $|t|<0.3$~{\gevsq}.
Since a  linear exponent is not
sufficient to describe the data at higher $|t|$,   the quadratic form  
exp$(-b|t|+ct^2)$ was also  fitted to both the BPC and DIS $\rho^0$ data 
for $|t|<0.6$~{\gevsq}. 
The results from the linear fits are 
$b=8.5\pm0.2$(stat.)$\pm0.5$(syst.)$\pm0.5$(pdiss.)~{\gevmsq} 
for  the  BPC $\rho^0$ sample ($0.25<Q^2<0.85$~{\gevsq} and  $20<W<90$~{\gev})
and
$b=8.1\pm0.6$(stat.)$\pm0.7$(syst.)$\pm0.7$(pdiss.)~{\gevmsq}
for the  DIS $\rho^0$ sample, which covers the kinematic region
depicted in Fig.~\ref{fig:kinematics}.
Since a  major  contribution to the systematic uncertainty 
arises from the uncertainty associated with 
subtracting the  proton  dissociation  background,
the error from this source is explicitly quoted.
In order to  illustrate the significance of the quadratic term in the
exponent of the fitted function, the  uncertainty due to the 
systematic error in the parameter $c$ is indicated by a shaded band 
in Fig.~\ref{fig:dNdt}.
Detailed results of fits in  $Q^2$,  $W$ and $M_{\pi\pi}$  intervals 
are summarised  in Tables~\ref{tab:tslope1} and  \ref{tab:tslope2}.

\begin{figure}
\centerline{
\psfig
{figure=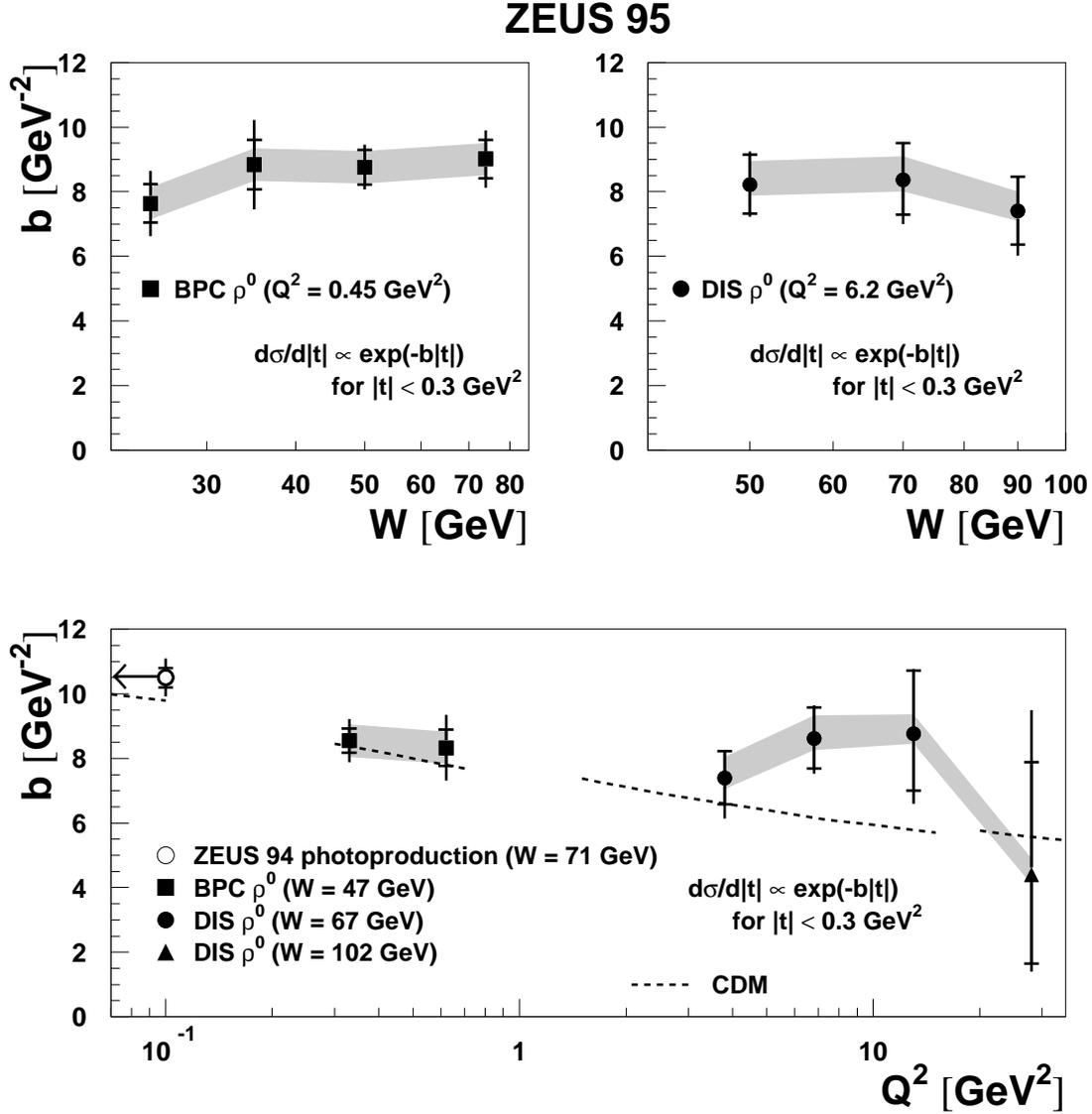,width=\textwidth}
}
\caption{
\label{fig:q2dep_b}
The slope parameter $b$ as a function of $W$ and $Q^2$ for 
exclusive $\pi^+\pi^-$  production (BPC and DIS  samples) in the
range $0.6 < M_{\pi\pi} < 1.2$~{\gev}. 
The open point with the horizontal arrow indicates 
the value measured in photoproduction~\protect{\cite{ref:rhopdissZEUS}}.
The inner error bars
represent statistical uncertainties; 
the outer error bars indicate the quadratic sum of
statistical and systematic  uncertainties.
The shaded areas indicate additional normalisation uncertainties due to
the proton dissociation background subtraction. 
The dashed lines   represent predictions of the Colour Dipole Model (CDM)
of Nemchik et al.~\protect{\cite{ref:cdm}}
at the corresponding $Q^2$ and $W$ values.
}
\end{figure}

The  slope parameter $b$ as a function of $W$ and $Q^2$   is displayed 
in Fig.~\ref{fig:q2dep_b}.  The results  for BPC $\rho^0$ are consistent with 
a slow rise with $W$. The DIS $\rho^0$ results are consistent with
no $W$ dependence. Both results show significantly shallower slopes than
that measured in photoproduction. The CDM calculation~\cite{ref:cdm} 
is shown for comparison. Its prediction of a decrease with $Q^2$ is in
reasonable agreement with the data.

The linear-exponent fit  for  the entire 
DIS $J/\psi$ sample  ($2<Q^2<40$~{\gevsq} and  $55<W<125$~{\gev}) 
in the range  $|t|<1$~{\gevsq} yielded 
$b=5.1\pm1.1$(stat.)$\pm0.7$(syst.)~{\gevmsq},
a result consistent with  the value of 
$4.6\pm0.4$(stat.)$^{+0.4}_{-0.6}$(syst.)~{\gevmsq}
obtained  in exclusive  $J/\psi$ photoproduction~\cite{ref:zeuselpsi}.

\subsection{Shrinkage of the diffractive peak}

Shrinkage of the diffractive peak was studied by
reweighting iteratively 
the energy and $b$ dependence in the Monte Carlo simulation according to
\begin{equation}
\frac{{\rm d}\sigma^{\rm ep}}{{\rm d}|t|} \propto {\rm exp}\left [ (- b_0 |t| + c_0 t^2) 
\cdot (W/W_0)^{4 [(\alpha(0)-1)- \alpha^{\prime} |t|]} \right ], 
\end{equation}
where $W_0$ is a constant,  $t$ and $W$ are the generated 
variables, and $b_0$, $\alpha(0)$ and $\alpha^{\prime}$ are the parameters 
tuned to the best agreement between the simulated and measured distributions.
The fit for the BPC data
yielded  $\alpha(0)$=1.055$\pm$0.016(stat.)$\pm$0.019(syst.)
and $\alpha^{\prime}$=0.19$\pm$0.09(stat.)$\pm$0.09(syst.)~GeV$^{-2}$, 
showing evidence for shrinkage, in  agreement with theoretical
predictions~\cite{ref:dl_vm}.
A similar analysis performed using  the DIS $\rho^0$ data  gave
an inconclusive result.

\subsection{Decay angular distributions}

\begin{figure}
\centerline{
\psfig
{figure=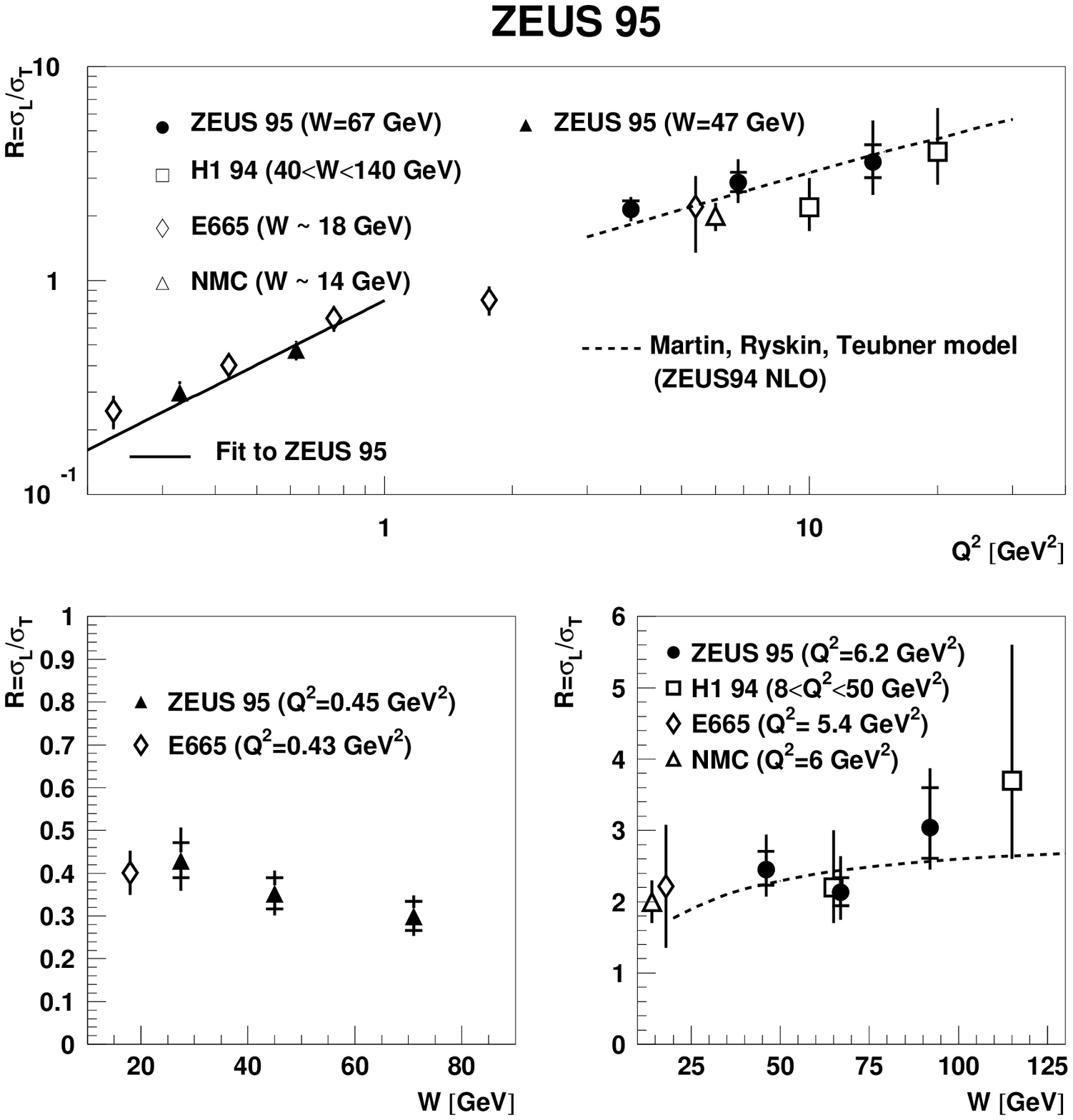,width=\textwidth}
}
\caption{
\label{fig:q2dep_R}
The ratio of the cross sections for longitudinal and transverse photons,
$R=\sigma_{\rm L}^{\gamma^{\star}{\rm p}} / \sigma_{\rm T}^{\gamma^{\star}{\rm p}}$,  for exclusive $\pi^+\pi^-$ production in the range 
$0.6<M_{\pi\pi}<1.2$~{\gev}
as a function of $Q^2$ and $W$, evaluated  assuming SCHC.
The inner error bars represent statistical
uncertainties; the outer error bars indicate the quadratic sum of
statistical and systematic  uncertainties.
The solid line represents the result of a fit to the BPC data of the form
$R=\kappa Q^2$, which yielded $\kappa=0.81\pm0.05$(stat.)$\pm0.06$(syst.).
The dashed line is a  prediction of the model 
by Martin, Ryskin and Teubner  \protect{\cite{ref:mrt}} using
the ZEUS 94 NLO  parameterisation of the gluon density~\protect{\cite{ref:zeus94nlo}}.
}
\end{figure}

The $\rho^0$ spin-density matrix elements,
$r^{04}_{00}$,  $r^1_{1-1}$ and Re$\:r^{5}_{10}$,
were determined by  a two-dimensional maximum-likelihood 
fit of Eq.~\ref{eq:angular}
to the $\cos\theta_{\rm h}$ and $\psi_{\rm h}$ distributions.
The results  are presented in Table~\ref{tab:matrix}.
The corresponding values of the ratio $R=\sigma_{\rm L}^{\gamma^{\star}{\rm p}} / \sigma_{\rm T}^{\gamma^{\star}{\rm p}}$  
as a function of $Q^2$ are displayed in the upper plot
of Fig.~\ref{fig:q2dep_R}. 
The ratio was evaluated according to  Eq.~\ref{eq:R_ratio}.
The results indicate that the ratio increases with $Q^2$, the $Q^2$ dependence
steeper at lower $Q^2$.
At high values of $Q^2$ the longitudinal cross section dominates.
The solid line represents the result of a fit to the BPC data of the form
$R=\kappa Q^2$, which yielded $\kappa=0.81\pm0.05$(stat.)$\pm0.06$(syst.).
The dashed line represents the results of
the QCD-based calculation of ref.~\cite{ref:mrt}, which describes the data
well. The lower plots of Fig.~\ref{fig:q2dep_R} show $R$ as a function of
$W$ for $Q^2=0.45$~{\gevsq} and 6.2~{\gevsq}. For $Q^2=0.45~{\gevsq}$, the data indicate a slow decrease of $R$ with $W$. However, they are 
consistent with no dependence
within two standard deviations. For $Q^2=6.2~{\gevsq}$, the measurements indicate
a slow rise of $R$ with $W$. 
The prediction of the model of ref.~\cite{ref:mrt} is in good agreement with these results.

The values of the spin-density matrix elements  $r^{04}_{00}$ and  $r^1_{1-1}$ 
satisfy  Eq.~\ref{eq:NPE}
within experimental uncertainties, and are thus 
consistent with natural-parity  exchange in the $t$ channel.

The $J/\psi$ spin-density matrix elements $r^{04}_{00}$ and $r^{04}_{1-1}$ 
were determined by one-dimensional fits of formulae~(\ref{eq:jpsi_angular1}) 
and~(\ref{eq:jpsi_angular2}).  The results for the entire  kinematic 
region covered by the data, for which \mbox{$<Q^2>=5.9$~{\gevsq}} and \mbox{$<W>=97$~\gev,} are
\begin{eqnarray*}
r^{04}_{00}\;=\;0.29\,\pm\,0.19\,{\rm (stat.)}\,^{+0.12}_{-0.18}\,{\rm (syst.)}, \hspace*{1cm}
r^{04}_{1-1}\;=\;-0.04\,\pm\,0.20\,{\rm (stat.)}\,^{+0.12}_{-0.22}\,{\rm (syst.)}. 
\end{eqnarray*}
Using \mbox{$<{\epsilon}>=0.99$,} a
value of $R$ of $0.41^{+0.45}_{-0.52}$ (statistical and systematic uncertainties
added in quadrature) was extracted, significantly less than the values
measured for the $\rho^0$ at similar $Q^2$.

\subsection{Forward longitudinal cross sections}

\begin{figure} 
\centerline{
\psfig
{figure=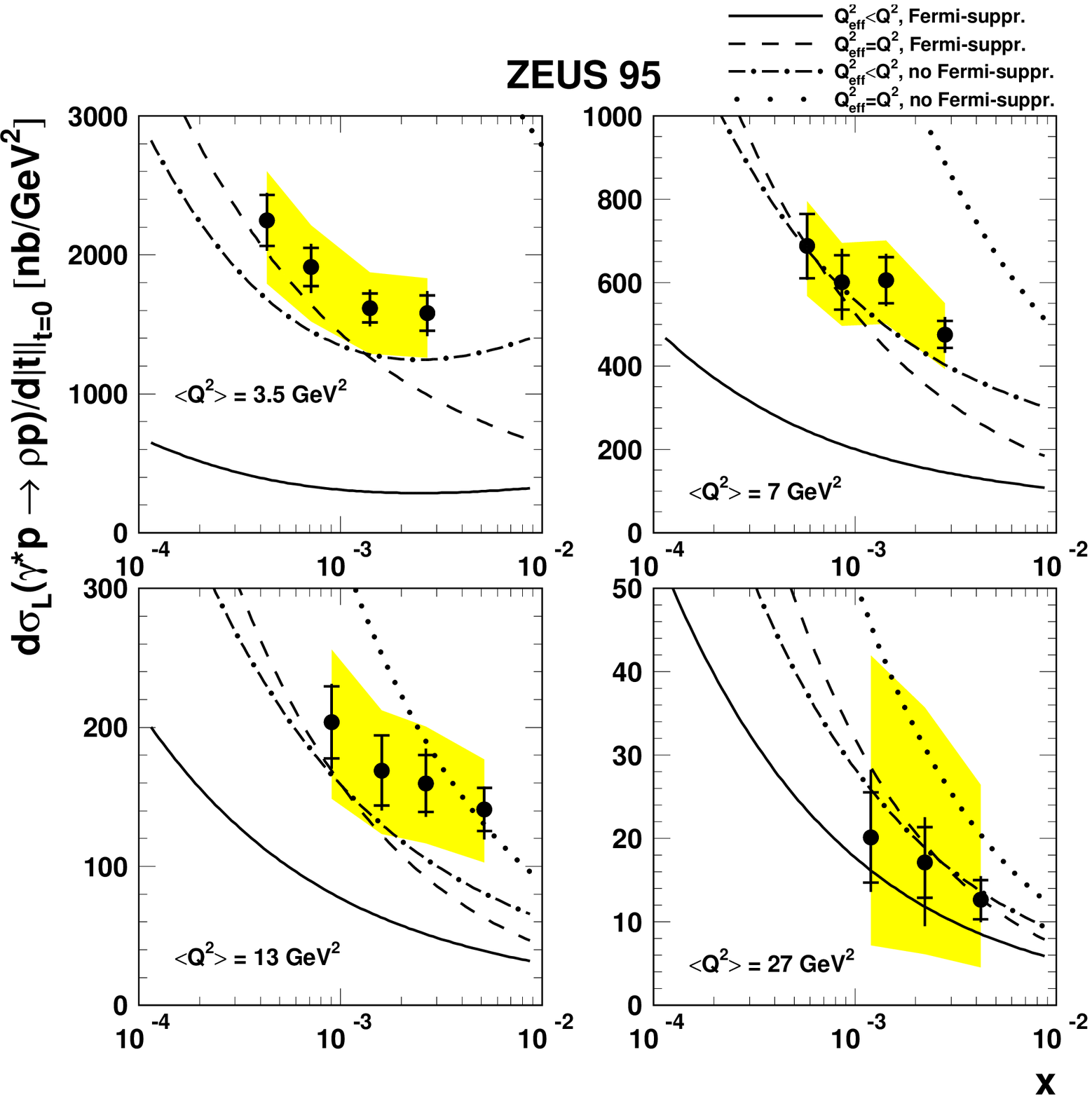,width=\textwidth}
}
\caption{
\label{fig:koepf}
The measured forward longitudinal cross section, $\left. {\rm d}\sigma_{\rm L}^{\gamma^{\star}{\rm p}} /{\rm d}|t| \right| _{t=0}$, 
as a function of $x$ for the DIS $\rho^0$ sample.
The inner error bars represent statistical
uncertainties; the outer error bars indicate the quadratic sum of
statistical and systematic  uncertainties.
The shaded areas indicate additional normalisation uncertainties due to
the proton dissociation background subtraction 
as well as the  measured values of the 
$R=\sigma_{\rm L}^{\gamma^{\star}{\rm p}} / \sigma_{\rm T}^{\gamma^{\star}{\rm p}}$ ratio  and the slope parameter $b$.
The curves show the predictions by Frankfurt, Koepf and
Strikman~\protect{\cite{ref:fks}}
using the ZEUS 94 NLO gluon parameterisation~\protect{\cite{ref:zeus94nlo}}.
The full and dashed lines show the result of the calculation 
assuming hard Fermi suppression
with rescaling ($Q^2_{{\rm eff}}<Q^2$) and without rescaling ($Q^2_{{\rm
eff}}=Q^2$). The dashed-dotted and dotted lines show the result assuming 
no hard Fermi suppression with and without rescaling.
}
\end{figure}

\begin{figure} 
\centerline{
\psfig
{figure=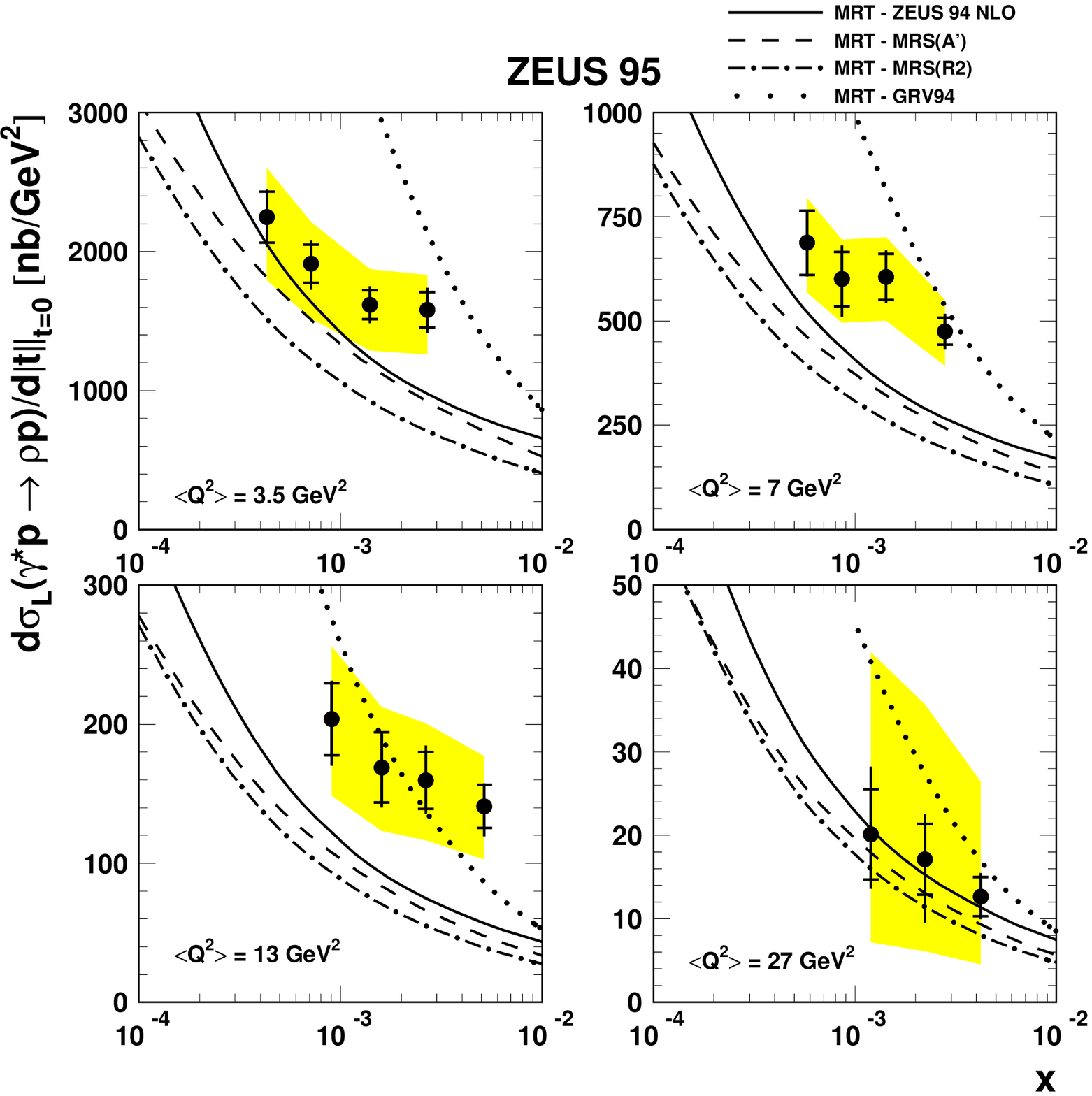,width=\textwidth}
}

\caption{
\label{fig:mrtsigl}
The measured forward longitudinal cross section, $\left. {\rm d}\sigma_{\rm L}^{\gamma^{\star}{\rm p}} /{\rm d}|t| \right| _{t=0}$, 
as a function of $x$
for the DIS $\rho^0$ sample.
The inner error bars represent statistical
uncertainties; the outer error bars indicate the quadratic sum of
statistical and systematic  uncertainties.
The shaded areas indicate additional normalisation uncertainties due to
the
proton dissociation background subtraction as well as the 
measured values of the 
$R=\sigma_{\rm L}^{\gamma^{\star}{\rm p}} / \sigma_{\rm T}^{\gamma^{\star}{\rm p}}$ ratio  and the $t$-slope parameter $b$.
The curves  show  the predictions by Martin, Ryskin and
Teubner~\protect{\cite{ref:mrt}} and correspond to various
gluon parameterisations, indicated as follows: full lines -- 
ZEUS 94 NLO~\protect{\cite{ref:zeus94nlo}},
dashed lines --  MRSA$^{\prime}$~\protect{\cite{ref:mrsaprime}}, dashed-dotted lines --  MRSR2~\protect{\cite{ref:mrsr2}}, and dotted
lines --  GRV94~\protect{\cite{ref:grv94}}.
}
\end{figure}

\begin{figure} 
\centerline{
\psfig{figure=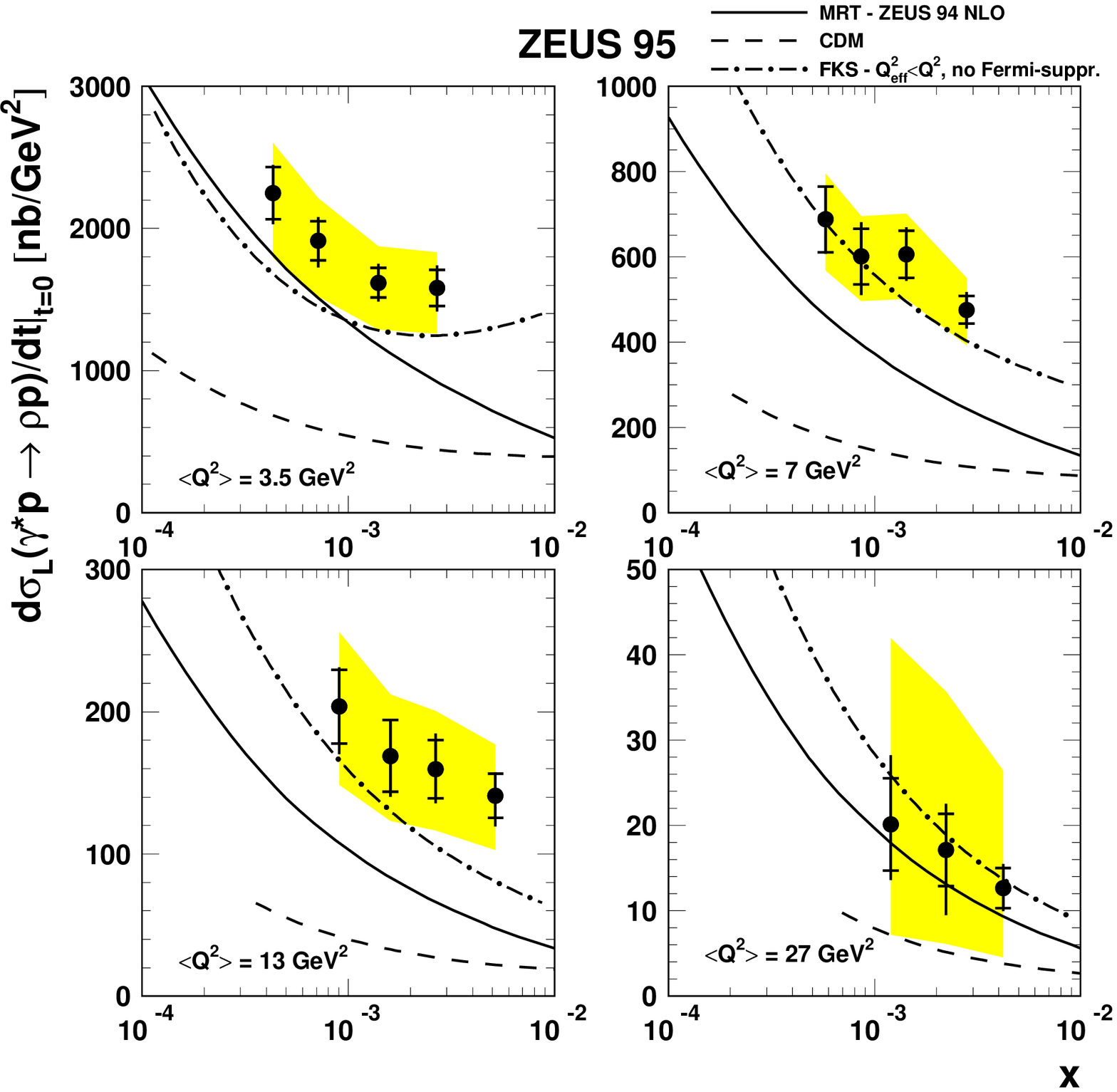,width=\textwidth}
}
\caption{
\label{fig:cdm}
The measured forward longitudinal cross section, $\left. {\rm d}\sigma_{\rm L}^{\gamma^{\star}{\rm p}} /{\rm d}|t| \right| _{t=0}$, 
as a function of $x$ for the DIS $\rho^0$ sample.
The inner error bars represent statistical
uncertainties; the outer error bars indicate the quadratic sum of
statistical and systematic  uncertainties.
The shaded areas indicate additional normalisation uncertainties due to
the proton dissociation background subtraction 
as well as the  measured values of the 
$R=\sigma_{\rm L}^{\gamma^{\star}{\rm p}} / \sigma_{\rm T}^{\gamma^{\star}{\rm p}}$ ratio  and the slope parameter $b$.
The solid line shows the calculation by Martin, Ryskin and 
Teubner~\protect{\cite{ref:mrt}} 
using the ZEUS 94 NLO gluon density parameterisation. The dashed line
shows the result of a calculation using CDM~\protect\cite{ref:cdm}.
The dashed-dotted line shows the prediction of Frankfurt, Koepf 
and Strikman~\protect{\cite{ref:fks}}
using  the ZEUS 94 NLO gluon parameterisation with rescaling 
($Q^2_{{\rm eff}}<Q^2$) and no hard Fermi suppression. 
}
\end{figure}

In order to compare our results to pQCD calculations, we extract the
forward longitudinal $\rho^0$ cross section according to
\begin{eqnarray}
\left.\frac{{\rm d}\sigma_{\rm L}^{\gamma^{\star}{\rm p}}}{{\rm d}|t|}\right|_{t=0} &=&\left(\frac{R}{1+R}\right)\;\left(\frac{b}{1-e^{-b\,|t|_{{\rm max}}}}\right)\;\sigma_{\rm tot}^{\gamma^{\star}{\rm p}},
\end{eqnarray}
where $|t|_{{\rm max}}$ is the upper limit on $|t|$ for which the cross section
was calculated.
A comparison of the measured  and the predicted  $x$ dependence of the
longitudinal 
$\rho^0$ production  cross section at  various values of 
$Q^2$ is shown in \mbox{Figs.~\ref{fig:koepf}-\ref{fig:cdm}.}  
The shaded areas indicate normalisation uncertainties due to
the proton dissociation background subtraction, the measured values of $R$  and of the slope parameter $b$. (For the highest $Q^2$ value, extrapolations of the $R$ and $b$ values were used.)

In the model of Frankfurt, Koepf and Strikman~\cite{ref:fks}
the hard diffractive production of vector mesons by longitudinal photons 
is calculated in the 
leading-order  approximation ($\alpha_{\rm s} \ln \frac{Q^2}{\Lambda^2}$)
using leading-order parton distributions.  
Rescaling effects are accounted for by introducing an effective scale,
$Q^2_{\rm eff}$, at which the gluon density is evaluated. The curves in Fig.~\ref{fig:koepf}, which use the ZEUS 94 next-to-leading-order (NLO) 
gluon density parameterisation~\cite{ref:zeus94nlo}, 
show the degree to
which the rescaling
attenuates the $Q^2$ dependence. Also shown are the effects of two
assumptions
concerning the $\rho^0$ wave function which result in different Fermi
motion suppression factors as calculated by the authors. 
The assumption of a hard Fermi suppression attenuates the $Q^2$ dependence, 
as does the rescaling, but with a different $x$ dependence. 
The two effects are of comparable magnitude
in the kinematic region covered by the data. 
The measurements indicate that the assumption of
a hard Fermi suppression together with the $Q^2$-rescaling results in an
overcorrection. Clearly a quantitative understanding of the higher-order 
QCD corrections is necessary before information on the
gluon density and on the $\rho$ meson wave function can be extracted.

The  model of Martin, Ryskin and Teubner~\cite{ref:mrt} is based
on the parton--hadron duality hypothesis, applied to the production
of {\qqbar} pairs. 
A comparison of the predictions using various gluon density functions
to the measured  $x$ dependence of the forward longitudinal 
 cross section is shown in Fig.~\ref{fig:mrtsigl}.  The 
MRSA$^{\prime}$~\cite{ref:mrsaprime},
MRSR2~\cite{ref:mrsr2}, and ZEUS 94 NLO gluon density parameterisations lead to similar predictions, whereas the prediction using the GRV94 parameterisation~\cite{ref:grv94} is considerably higher. In the context of this model, the data are sufficiently precise to distinguish between GRV94 and the other parton density functions.

Fig.~\ref{fig:cdm} compares the calculations of the two models described
above with that of Nemchik et al.~\cite{ref:cdm}, which is based on colour
dipole BFKL phenomenology. Here, the model of Frankfurt, Koepf, and Strikman
uses the 
ZEUS 94 NLO gluon density function with rescaling and no hard Fermi suppression.
The curves for the model of Martin, Ryskin, and Teubner represent a calculation
which also employs the ZEUS
94 NLO gluon density parameterisation. With these choices, the models describe the data reasonably well, taking into account the normalisation uncertainties.
Note that the normalisation uncertainty due to the uncertainty in $b$ is 
largely independent for the various $Q^2$ values and dominates at high $Q^2$.
The model of Nemchik et al. underestimates the cross
section over the entire kinematic range investigated.

%
%

\section{Summary and conclusions}

We have studied the exclusive  electroproduction 
of $\rho^0$, ${\rm ep}\rightarrow {\rm e}\rho^0 {\rm p}$, and $J/\psi$ mesons,
${\rm ep}\rightarrow {\rm e} J/\psi\; {\rm p}$,
in the kinematic range 
$0.25 < Q^2 < 50$~{\gevsq}, $20 < W < 167$~{\gev} for the $\rho^0$ data 
and $2 < Q^2 < 40$~{\gevsq}, $50 < W < 150$~{\gev} for the $J/\psi$ data.
The results can be summarised as follows.

\begin{itemize}

\item 

The  $\pi^+\pi^-$ mass spectrum for exclusively produced
$\rho^0$ mesons shows 
a deviation  from the  relativistic p-wave Breit-Wigner shape.
This can be explained in terms of the  interference between resonant
and nonresonant production amplitudes.
The relative contribution of the nonresonant amplitude is found to
decrease with  $Q^2$ and becomes consistent with zero  at $Q^2\simeq$20~{\gevsq}.

\item

The $Q^2$ dependence  of the $\gamma^{\star} {\rm p} \rightarrow \rho^0 p$ 
cross section at  low $Q^2$ ($0.25<Q^2<0.85$~{\gevsq})
can be described by the function  
$\sigma_{\rm tot}^{\gamma^{\star}{\rm p}}  \propto 
1/(1+Q^2/M_{\rho}^2)^n$ with $n=1.75\pm0.10$ (stat.)$\pm$0.29(syst.).
At higher values of $Q^2$    the dependence  can be fitted with 
the function $\sigma_{\rm tot}^{\gamma^{\star}{\rm p}} 
\propto Q^{-2n}$ with  the average fitted
value of $n=2.3\pm0.1$(stat.), essentially independent of $W$.
For the DIS $J/\psi$ sample the data  are  described by the function 
$\sigma_{\rm tot}^{\gamma^{\star}{\rm p}} \propto 1/(1+Q^2/M_{J/\psi}^2)^n$, 
with $n$=1.58$\pm$0.22(stat.)$\pm$0.09(syst.).

\item 

The $W$ dependence of the $\gamma^{\star} {\rm p} \rightarrow \rho^0 p $  cross section
exhibits a slow rise with $W$ at low values of $Q^2$.
Parameterising the cross section as   
$\sigma_{\rm tot}^{\gamma^{\star}{\rm p}} \propto W^{\delta}$
yields the fit result $\delta$=0.12$\pm$0.03(stat.)$\pm$0.08(syst.)
for $Q^2$=0.47~{\gevsq} (BPC $\rho^0$).
This value is consistent with that measured in  photoproduction
as well as with predictions based on  soft pomeron exchange~\cite{ref:dl}.  
The slope becomes steeper with increasing  $Q^2$.
For $3.5 < Q^2 < 13$~{\gevsq}  the average value is $0.42\pm0.12$.
This is less steep than the value of
$\delta$=0.92$\pm$0.14(stat.)$\pm$0.10(syst.)
measured  in $J/\psi$ photoproduction~\cite{ref:zeuselpsi}.
The cross section for $J/\psi$
electroproduction has a $W$ dependence consistent with the steep dependence
found in photoproduction.

\item

The ratio $\sigma(J/\psi)/\sigma(\rho^{0})$ 
increases with $Q^2$ but does not reach the flavour-symmetric expectation 
of 8/9 at $Q^2=13$~{\gevsq}.

\item
The $t$ distributions for  exclusive $\rho^0$ production
are well described by an exponential 
dependence ${\rm d}\sigma^{{\rm ep}}/{\rm d}|t|\propto e^{-b|t|}$ for $|t|<0.3$~{\gevsq}
with  $b\simeq 8$~{\gevmsq}.
The slope decreases  at larger values of $|t|$. The Colour Dipole Model~\cite{ref:cdm} gives a reasonable description of the data.
A lower  value, $5.1\pm1.1$(stat.)$\pm0.7$(syst.)~{\gevmsq}, 
has been obtained  in  exclusive $J/\psi$ production in the kinematic
region $2<Q^2<40$~{\gevsq} for $|t|<1$~{\gevsq}. 
This result is compatible with that
for $J/\psi$ photoproduction. 

\item
The $\rho^0$ measurements in the range $0.25<Q^2<0.85$~{\gevsq} 
exhibit a $W$ dependence 
in the $|t|$ distribution. This may be interpreted  as due to the  shrinkage
of the diffractive peak, predicted in Regge theory. In this context,
we find $\alpha(0)$=1.055$\pm$0.016(stat.)$\pm$0.019(syst.)
and $\alpha^{\prime}$=0.19$\pm$0.09(stat.)$\pm$0.09(syst.)~GeV$^{-2}$.
Tests for shrinkage in the DIS $\rho^0$ sample were inconclusive.

\item
The ratio of the cross sections for longitudinal and transverse photons,
$R=\sigma_{\rm L}^{\gamma^{\star}{\rm p}} / \sigma_{\rm T}^{\gamma^{\star}{\rm p}}$,  increases  with $Q^2$ and shows a weak $W$ dependence.    
For $Q^2>3${~\gevsq} these dependences are well reproduced  
by the model of Martin, Ryskin and 
Teubner~\cite{ref:mrt}. For $J/\psi$ electroproduction, $R \simeq 0.4$ at 
$Q^2=6$~{\gevsq}, in contrast to the value of $R \gsim 2$ for the $\rho^0$ meson.

\item

The measurements of the forward longitudinal cross section, 
${\rm d}\sigma_{\rm L}^{\gamma^{\star}{\rm p}}/{\rm d}|t||_{t=0}$, 
for $\rho^0$ production  have been compared to the results of
calculations based on several  pQCD models.
The present level of accuracy in the measurements allows
quantitative distinctions between the various calculations.

\end{itemize}

In conclusion, our results for exclusive $\rho^0$ production
show the $Q^2$ range $0.25<Q^2<50$~{\gevsq} to be a transition region, where,
as $Q^2$ increases, the relative contribution of continuum $\pi^+\pi^-$
production decreases, and the longitudinal contribution to the total
cross section increases and becomes dominant. These trends encourage efforts
to describe this process using the methods of perturbative QCD.

Exclusive $J/\psi$ electroproduction is consistent with 
expectations from pQCD.   The exponential slope of the $|t|$ dependence is 
approximately $5$~{\gevmsq} and the $W$ dependence of the cross section is 
consistent with the steep rise observed in $J/\psi$ photoproduction.  
These dependences differ from those measured in $\rho^0$ 
electroproduction at $Q^2\simeq M_{J/\psi}^2$, where
the $|t|$ dependence is  steeper and the $W$ dependence shallower.  
We also find contrasting values for $R$ in $J/\psi$ and $\rho^0$ 
electroproduction. Thus $Q^2$ and $M^2_{\rm V}$ are 
shown to play dissimilar roles in
setting the scale of the process.


\section*{Acknowledgements}
We thank the DESY Directorate for their strong support and encouragement,
and the HERA machine group for their diligent efforts.
We are grateful for the  support of the DESY computing
and network services. The design, construction and installation
of the ZEUS detector have been made possible owing to the ingenuity and effort
of many people from DESY and home institutes who are not listed
as authors. It is also a pleasure to thank L.~Frankfurt, W.~Koepf, A.D.~Martin, J.~Nemchik, N.N.~Nikolaev, M.G.~Ryskin, M.~Strikman 
and T.~Teubner for many useful discussions and for providing
model calculations.

\newpage

\begin{table}
\begin{center}
\begin{tabular}{|l|c|c|c|}               \hline \hline
&&&\\
 Contribution from                     &  BPC $\rho^0$ & DIS $\rho^0$ & DIS $J/\psi$ \\ 
&&&\\
\hline 
 Luminosity                             &   1.1*         & 1.1*          & 1.1*      \\ 
 Acceptance: trigger efficiency         &   5.5*         & $<$1*         & $<$1*     \\ 
 Acceptance: model dependence           &   1-4         & 5*            &   1      \\ 
 Acceptance: electron identification    &   3-10        & $<$1         & $<$1     \\ 
 Acceptance: dependence on  cuts        &   2-10        & 6      & $^{+6}_{-11}$   \\
 Acceptance: photon flux determination  &   1*           & 1            &   1*    \\ 
 Procedure to extract the signal events &   10*          &1*             &   5*      \\ 
 Proton diss. background subtraction    &   10*           & $^{+7}_{-12}$*& $^{+11}_{-13}$*   \\ 
 Elastic $\omega$ and $\phi$ production &  1.6*          & $<$1*         &   --     \\ 
 Elastic $\psi^{\prime}$ production     &     --        &      --      &   1*       \\ 
 Photon diffractive dissociation        &$^{+0}_{-3}$*  & $^{+0}_{-3}$* &   --       \\ 
 Radiative corrections                  &    2*          &  1*           &   4*        \\ 
 Beam-gas interactions                  &$^{+0}_{-1.5}$* &$^{+0}_{-1.5}$* &$^{+0}_{-1.5}$*\\ 
 Branching ratio                        &   --          &   --         &   2.2*      \\\hline \hline
\end{tabular}
\caption{
\label{syserr}
Typical values of relative contributions (\%) to the systematic  
uncertainty  in the integrated cross sections
presented in  Tables~\protect{\ref{tab:xs_rho}} and~\protect{\ref{tab:xs_jpsi}}.
The starred values are the contributions to the overall normalisation
uncertainty in each of the three analyses.
}
\end{center}
\end{table}

\clearpage


\newpage

\begin{table}[t]
\begin{center}
\begin{small}
\begin{tabular}{|c|c|r|c|c|c|c|}\hline
$W$ & $Q^2$ &\#evts & $W_0$ & $Q^2_0$ & $\sigma^{\rm ep}$ & $\sigma^{\gamma^{\star}{\rm p}}$\\
 $[{\gev}]$ & $[{\gevsq}]$ & & $[{\gev}]$ & $[{\gevsq}]$ & $[nb]$ & [$\mu$b]\\
\hline
\multicolumn{7}{|c|}{{BPC $\rho^0$}}\\
\hline
      & 0.25-0.29 & 1074 &      & 0.27 & 4.99$\pm$0.15$\pm$0.47 &  5.07$\pm$0.15$\pm$0.48 \\
      & 0.29-0.33 &  941 &      & 0.31 & 3.98$\pm$0.14$\pm$0.47 &  4.64$\pm$0.16$\pm$0.55 \\
20-90 & 0.33-0.38 &  857 & 51.1 & 0.35 & 3.88$\pm$0.13$\pm$0.33 &  4.14$\pm$0.14$\pm$0.36 \\
      & 0.38-0.45 &  869 &      & 0.41 & 4.34$\pm$0.15$\pm$0.56 &  3.86$\pm$0.13$\pm$0.49 \\
      & 0.45-0.55 &  784 &      & 0.50 & 4.55$\pm$0.16$\pm$0.49 &  3.41$\pm$0.12$\pm$0.37 \\
      & 0.55-0.85 &  937 &      & 0.69 & 7.28$\pm$0.23$\pm$1.06 &  2.51$\pm$0.08$\pm$0.37 \\
\hline
20-27 &           &  955 & 23.4 &      & 5.52$\pm$0.18$\pm$0.64 &  3.16$\pm$0.10$\pm$0.37 \\
27-35 &           & 1024 & 30.9 &      & 4.75$\pm$0.15$\pm$0.59 &  3.16$\pm$0.10$\pm$0.40 \\
35-45 & 0.25-0.85 &  994 & 39.9 & 0.47 & 4.62$\pm$0.24$\pm$0.61 &  3.19$\pm$0.16$\pm$0.43 \\
45-55 &           &  897 & 49.9 &      & 4.28$\pm$0.15$\pm$0.56 &  3.74$\pm$0.13$\pm$0.48 \\
55-70 &           & 1018 & 62.4 &      & 4.89$\pm$0.15$\pm$0.50 &  3.61$\pm$0.11$\pm$0.38 \\
70-90 &           &  574 & 79.8 &      & 4.72$\pm$0.20$\pm$0.56 &  3.44$\pm$0.14$\pm$0.40 \\
\hline
\multicolumn{7}{|c|}{{DIS $\rho^0$}}\\
\hline
 32--40& 3--5&254& 36&3.5& $0.141 \pm 0.011 ^{+0.009}_{-0.010}$ & $0.310 \pm 0.025 ^{+0.019}_{-0.022}$\\
 40--60&     &492& 50&   & $0.256 \pm 0.017 ^{+0.014}_{-0.014}$ & $0.318 \pm 0.021 ^{+0.017}_{-0.017}$\\
 60--80&     &401& 70&   & $0.211 \pm 0.016 ^{+0.011}_{-0.014}$ & $0.376 \pm 0.027 ^{+0.019}_{-0.025}$\\
 80--100&     &318& 90&   & $0.186 \pm 0.016 ^{+0.006}_{-0.014}$ & $0.443 \pm 0.036 ^{+0.014}_{-0.025}$\\
\hline
 40--60& 5--10&380& 50&  7 & $0.101 \pm 0.007 ^{+0.006}_{-0.006}$ & $0.075 \pm 0.005 ^{+0.004 }_{-0.005 }$\\
 60--80&      &331& 70&    & $0.089 \pm 0.008 ^{+0.004}_{-0.004}$ & $0.095 \pm 0.009 ^{+0.005 }_{-0.005 }$\\
 80--100&      &234& 90&   & $0.066 \pm 0.007 ^{+0.005}_{-0.007}$ & $0.094 \pm 0.010 ^{+0.007 }_{-0.010 }$\\
100--120&      &193&110&   & $0.058 \pm 0.006 ^{+0.001}_{-0.002}$ & $0.109 \pm 0.012 ^{+0.002 }_{-0.004 }$\\
\hline
 41--60&10--20&106& 50& 13& $0.023 \pm 0.002 ^{+0.001}_{-0.002}$ & $0.021 \pm 0.002 ^{+0.001 }_{-0.002 }$\\
 60--80&     & 88& 70&    & $0.019 \pm 0.002 ^{+0.002}_{-0.002}$ & $0.024 \pm 0.003 ^{+0.002 }_{-0.002 }$\\
 80--100&     & 72& 90&   & $0.014 \pm 0.002 ^{+0.001}_{-0.002}$ & $0.025 \pm 0.004 ^{+0.001 }_{-0.002 }$\\
100--140&     &110&120&   & $0.025 \pm 0.003 ^{+0.001}_{-0.003}$ & $0.030 \pm 0.004 ^{+0.001 }_{-0.003 }$\\
\hline
 55--96&20--50& 27& 80& 27 & $0.006 \pm 0.001 ^{+0.001}_{-0.001}$ & $0.0033 \pm 0.0007^{+0.0004}_{-0.0004}$\\
 96--125&      & 17&110&   & $0.004 \pm 0.001 ^{+0.001}_{-0.002}$ & $0.0045 \pm 0.0012^{+0.0010}_{-0.0018}$\\
125--167&      & 16&150&   & $0.004 \pm 0.001 ^{+0.001}_{-0.001}$ & $0.0053 \pm 0.0015^{+0.0017}_{-0.0010}$\\
\hline\hline
\end{tabular}
\end{small}
\caption{
\label{tab:xs_rho}
Exclusive $\rho^0$ production cross sections 
for $|t|<0.6~{\gevsq}$ in 
various $Q^2$ and $W$ intervals.
The BPC $\rho^0$ cross sections are calculated 
for the invariant  mass range 
$2m_{\pi} < M_{\pi\pi} < M_{\rho} + 5 \Gamma_{\rho}$.
The cross sections are  given at $Q^2_0$ and $W_0$ values
assuming the $Q^2$ and $W$ dependence from this analysis. 
The uncertainties do not include the normalisation uncertainties,
which are
$^{+9\%}_{-14\%}$ for the 
DIS $\rho^0$ sample and, $^{+15\%}_{-16\%}$  for the BPC $\rho^0$
sample.
}
\end{center}
\end{table}

\clearpage

\newpage

\begin{table}[htbp]
\begin{center}
\begin{tabular}{|c|c|c|r|c|c|c|}\hline
$W$ [{\gev}]&$Q^2$ [{\gevsq}]&\# events&$W_0$ [{\gev}]&$Q^2_0$ [{\gevsq}]&$\sigma^{\rm ep}$ [pb]&$\sigma^{\gamma^{\star} {\rm p}}$ [nb]\\
\hline\hline
 50--100 & 2--7 & 31$\pm$7 &  70 & 3.5 &  79$\pm$18$^{+11}_{-12}$ & $21 \pm 5  \pm 3 $\\
100--150 & 2--7 & 20$\pm$7 & 120 & 3.5 &  56$\pm$19$^{+10}_{-11}$ & $29 \pm 10 \pm 6 $\\
 50--150 & 2--7 &          &  90 & 3.5 & 135$\pm$26$^{+18}_{-20}$ & $25 \pm 5  \pm 4 $\\
\hline
 50--100 &7--40 & 24$\pm$6 &  70 &13.0 &  30$\pm$7$^{+4}_{-5}$   & $6  \pm 2 \pm 1 $\\
100--150 &7--40 & 29$\pm$7 & 120 &13.0 &  39$\pm$9$^{+5}_{-6}$   & $17 \pm 4 \pm 3 $\\
 50--150 &7--40 &          &  90 &13.0 &  69$\pm$12$^{+8}_{-10}$ & $10 \pm 2 \pm 2 $\\
\hline
\end{tabular}
\caption{
\label{tab:xs_jpsi}
Exclusive $J/\psi$ production cross sections
in 
various $Q^2$ and $W$ intervals.
Values  are quoted  at  $Q^2_0$ and $W_0$.
The first error is statistical, the  second  systematic.
The systematic uncertainties include the normalisation uncertainty of $^{+13\%}_{-15\%}$ added in quadrature.
}
\end{center}
\end{table}

\begin{table}[htbp]
\begin{center}
\begin{tabular}{|c|c|} 
\hline
 $Q_0^2$ [{\gevsq}] &$\delta$\\  
\hline\hline
&   \\
0.47 &$0.12 \pm 0.03\pm 0.08$\\
&  \\                    
3.5 &$0.40 \pm 0.12\pm 0.12$\\
& \\
7.0 &$0.45 \pm 0.15\pm 0.07$\\
& \\
13.0 &$0.41 \pm 0.19\pm 0.10$\\
& \\
27.0 &$0.76 \pm 0.55\pm 0.60$\\
\hline\hline
\end{tabular}
\caption{
\label{tab:delta}
The values of the parameter $\delta$ obtained
by fitting the $W$ dependence of $\sigma_{\rm tot}^{\gamma^{\star}{\rm p}}$
for exclusive $\rho^0$ production with a  function  
$\sigma_{\rm tot}^{\gamma^{\star}{\rm p}} \propto  W^{\delta}$.
The first error is statistical, the  second  systematic.
}
\end{center}
\end{table}

\begin{table}[htbp]
\begin{center}
\begin{tabular}{|c|c|c|} 
\hline
$Q_{0}^{2}$ [{\gevsq}]  &  $W_{0}$ [{\gev}]  &  $\sigma(J/\psi)/\sigma(\rho^{0})$  \\ 
\hline\hline
          3.5          &        90       &  $0.06\pm0.01\pm0.01$  \\ 
         13.0          &        90       &  $0.43\pm0.10^{+0.03}_{-0.06}$  \\ 
\hline
\end{tabular}
\caption{
\label{Tab:RATIOjpsirho}
The ratio of  $J/\psi$ and \protect$\rho^{0}$ cross sections 
measured at two $Q^2$ values. The first error is statistical, the second
systematic. }
\end{center}
\end{table}

\clearpage

\newpage

\begin{table}[htbp]
\begin{center}
\begin{tabular}{|c|c|c|c|}\hline
&&&\\
$W_0$ [{\gev}]&$Q_0^2$ [{\gevsq}]&$M_{\pi\pi}$ [{\gev}]& $b$ [{\gevmsq}]\\
&&&\\
\hline\hline
\multicolumn{4}{|c|}{{BPC $\rho^0$}}\\
\hline
&&&\\
  47   & 0.45 &  $0.6<M_{\pi\pi}<1.2$ & $8.5 \pm 0.2 \pm 0.5 \pm 0.5$ \\*[3mm]
  25   &      &                       & $7.6 \pm 0.6 \pm 0.8 \pm 0.5$ \\
  35   &      &                       & $8.8 \pm 0.8 \pm 1.2 \pm 0.5$ \\
  50   &      &                       & $8.8 \pm 0.5 \pm 0.4 \pm 0.4$ \\
  74   &      &                       & $9.0 \pm 0.6 \pm 0.7 \pm 0.7$ \\*[3mm]
  47   & 0.33 &                       & $8.6 \pm 0.4 \pm 0.6 \pm 0.5$ \\
  47   & 0.62 &                       & $8.3 \pm 0.6 \pm 0.9 \pm 0.5$ \\*[3mm]
  47   & 0.45 &  $0.6<M_{\pi\pi}<0.7$ & $9.7 \pm 0.7 \pm 0.8 \pm 0.5$ \\
  47   &      &  $0.7<M_{\pi\pi}<0.8$ & $8.5 \pm 0.5 \pm 0.6 \pm 0.5$ \\
  47   &      &  $0.8<M_{\pi\pi}<1.2$ & $7.8 \pm 0.6 \pm 0.6 \pm 0.5$ \\

&&&\\
\hline
\multicolumn{4}{|c|}{{DIS $\rho^0$}}\\
\hline
&&&\\
 67  &  6.2 &$0.6<M_{\pi\pi}<1.2$& $ 8.1^{+0.6}_{-0.6}$$^{+0.3}_{-0.7}$$^{+0.7}_{-0.4}$ \\*[3mm]
 50  &  6.2 && $ 8.2^{+0.9}_{-0.9}$ $^{+0.4}_{-0.4}$$^{+0.7}_{-0.4}$ \\
 70  &  6.2 && $ 8.4^{+1.1}_{-1.1}$ $^{+0.3}_{-0.8}$$^{+0.7}_{-0.4}$ \\
 90  &  6.2 && $ 7.4^{+1.1}_{-1.0}$ $^{+0.2}_{-0.9}$$^{+0.6}_{-0.3}$ \\*[3mm]
 67  &  3.8 && $ 7.4^{+0.8}_{-0.8}$ $^{+0.1}_{-1.0}$$^{+0.7}_{-0.3}$ \\
 67  &  6.8 && $ 8.6^{+1.0}_{-0.9}$ $^{+0.4}_{-0.6}$$^{+0.7}_{-0.4}$ \\
 67  &  13  && $ 8.7^{+2.0}_{-1.8}$ $^{+0.5}_{-1.3}$$^{+0.6}_{-0.3}$ \\
102  &  28  && $ 4.4^{+3.5}_{-2.8}$ $^{+3.7}_{-1.2}$$^{+0.5}_{-0.3}$ \\*[3mm]
 67  &  6.2 &$0.6<M_{\pi\pi}<0.8$& $ 8.3^{+0.8}_{-0.7}$$^{+0.2}_{-0.5}$ $^{+0.8}_{-0.4}$ \\
 67  &  6.2 &$0.8<M_{\pi\pi}<1.2$& $ 7.7^{+1.0}_{-0.9}$$^{+0.7}_{-1.1}$ $^{+0.6}_{-0.3}$ \\
&&&\\
\hline\hline
\end{tabular}
\caption{
\label{tab:tslope1}
The values of the slope parameter $b$ 
obtained by fitting  ${\rm d}\sigma^{\rm ep}/{\rm d}|t| \propto e^{-b|t|}$
in the range $|t|<0.3$~{\gevsq} in various  $Q^2$, $W$, and $M_{\pi^+\pi^-}$   
ranges of the BPC and DIS $\rho^0$ samples. The first line of each of
the BPC and DIS sections indicates the results of the fit to the full
sample.
The first error is statistical, the 
second systematic and the third is the uncertainty
resulting from the subtraction
of the proton dissociation background.
}
\end{center}
\end{table}

\clearpage

\newpage

\begin{table}[t]
\begin{center}
\begin{tabular}{|c|c|c|c|}\hline
&&&\\
$W_0$ [{\gev}]&$Q_0^2$ [{\gevsq}]& $b$ [{\gevmsq}] & $c$ [GeV$^{-4}$] \\
&&&\\
\hline\hline
&&&\\
 47    & 0.45  & $9.5\pm0.3\pm0.6\pm0.5$ &  $4.0\pm0.7\pm0.8\pm0.4$   \\
&&&\\
\hline
&&&\\
 67  &  6.2 & $ 9.5\pm0.8\pm1.1\pm0.9$
& 6.1$\pm 1.3\pm1.7\pm0.5$  \\
&&&\\
\hline\hline
\end{tabular}
\caption{
\label{tab:tslope2}
The values of the  parameters $b$ and $c$ 
obtained by fitting  
$\sigma^{{\rm e p}}/{\rm d}|t| \propto e^{-b|t|+ct^2}$
in the range $|t|<0.6$~{\gevsq} for the full BPC and DIS $\rho^0$ samples. 
The mass range $0.6<M_{\pi^+\pi^-}<1.2~\gev$ was used.
The first error is statistical, the 
second systematic and the third is the uncertainty
resulting from the subtraction
of the proton dissociation background.
}
\end{center}
\end{table}

\clearpage

\begin{table}[t]
\begin{center}
\begin{tabular}{|c|c|c|}\hline
$W_0$ [{\gev}]&$Q_0^2$ [{\gevsq}]& $r^{04}_{00}$\\
\hline\hline
 27.5&  0.45& $ 0.30\pm0.02\pm0.03$\\ 
 45  &  0.45& $ 0.26\pm0.02\pm0.02$\\ 
 71  &  0.45& $ 0.23\pm0.02\pm0.02$\\ 
 47  &  0.33& $ 0.23\pm0.01\pm0.02$\\ 
 47  &  0.62& $ 0.32\pm0.01\pm0.02$\\ 
 46  &  6.2 & $ 0.71\pm0.02\pm0.03$\\          
 67  &  6.2 & $ 0.68\pm0.02\pm0.04$\\
 92  &  6.2 & $ 0.75\pm0.03\pm0.03$\\
\hline
 67  &  3.8 & $ 0.68\pm0.02\pm0.02$\\
 67  &  6.8 & $ 0.74\pm0.02\pm0.04$\\
 67  & 14.1 & $ 0.76\pm0.03\pm0.06$\\
\hline\hline
$W_0$ [{\gev}]&$Q_0^2$ [{\gevsq}]& $r^{1}_{1-1}$\\
\hline\hline
 27.5&  0.45& $ 0.32\pm0.01\pm0.03$\\ 
 45  &  0.45& $ 0.35\pm0.01\pm0.02$\\ 
 71  &  0.45& $ 0.36\pm0.01\pm0.02$\\ 
 47  &  0.33& $ 0.36\pm0.01\pm0.02$\\ 
 47  &  0.62& $ 0.31\pm0.01\pm0.03$\\ 
 46  &  6.2 & $ 0.12\pm0.02\pm0.02$\\                     
 67  &  6.2 & $ 0.15\pm0.02\pm0.06$\\
 92  &  6.2 & $ 0.10\pm0.02\pm0.04$\\
\hline
 67  &  3.8 & $ 0.15\pm0.02\pm0.02$\\
 67  &  6.8 & $ 0.11\pm0.02\pm0.02$\\
 67  & 14.1 & $ 0.08\pm0.03\pm0.07$\\
\hline\hline
$W_0$ [{\gev}]&$Q_0^2$ [{\gevsq}]& Re $r^{5}_{10}$\\
\hline\hline
 27.5&  0.45& $ 0.150\pm0.005\pm0.010$\\ 
 45  &  0.45& $ 0.136\pm0.004\pm0.010$\\ 
 71  &  0.45& $ 0.132\pm0.005\pm0.010$\\ 
 47  &  0.33& $ 0.133\pm0.004\pm0.010$\\ 
 47  &  0.62& $ 0.156\pm0.004\pm0.010$\\ 
 46  &  6.2 & $ 0.10\pm0.02\pm0.03$\\
 67  &  6.2 & $ 0.11\pm0.02\pm0.03$\\
 92  &  6.2 & $ 0.11\pm0.02\pm0.03$\\
\hline
 67  &  3.8 & $ 0.11\pm0.02\pm0.04$\\
 67  &  6.8 & $ 0.10\pm0.02\pm0.03$\\
 67  & 14.1 & $ 0.09\pm0.03\pm0.03$\\
\hline\hline
\end{tabular}
\caption{
\label{tab:matrix}
The spin-density matrix elements
$r^{04}_{00}$,  $r^{1}_{1-1}$ and Re $r^{5}_{10}$  
determined  using Eq.~\protect{\ref{eq:angular}}
for various values of $W$ and $Q^2$ (BPC and DIS $\rho^0$  samples). 
The first error is statistical, the second systematic.  
}
\end{center}
\end{table}

\clearpage

\clearpage

\newpage

\clearpage

\newpage

\clearpage

\newpage

\clearpage

\newpage

\clearpage


\newpage

\clearpage

\newpage

\clearpage

\newpage

\clearpage

%
%
%
\newpage

%
%
%
\newpage

\clearpage

%
%
%
\newpage
%
%
\newpage
%
%
\newpage
%
%
\newpage

\newpage


\newpage


\newpage

\newpage

\clearpage

\newpage

\clearpage

\newpage

\clearpage

\begin{thebibliography}{99}
\addcontentsline{toc}{chapter}{Bibliography}

\bibitem{ref:crittenden}
J.A.~Crittenden, {\em Exclusive Production of Neutral Vector Mesons at
the
  Electron-Proton Collider HERA,} {\rm Springer Tracts in Modern
Physics},
  Volume~140 (Springer, Berlin Heidelberg, 1997).
\bibitem{ref:bauer} 
R.M. Egloff et al., Phys. Rev. Lett. 43 (1979)~657; \\
R.M. Egloff et al., Phys. Rev. Lett. 43 (1979)~1545; \\
D. Aston et al.,  Nucl. Phys. B209 (1982)~56;\\
J. Busenitz et al., Phys. Rev. D40 (1989)~1.\\
For a review of earlier results, see T.H.~Bauer, R.D.~Spital, D.R.~Yennie and
F.M.~Pipkin, Rev. Mod. Phys.~50 (1978)~261; Erratum ibid. 51 (1979) 407; and references therein.
\bibitem{ref:rhoZEUS93} 
ZEUS Collab., M. Derrick et al., Z.Phys. C69 (1995)~39.
\bibitem{ref:jetphi}
ZEUS Collab., M.Derrick et al., Phys. Lett. B377 (1996)~259.
\bibitem{ref:heravm} 
ZEUS Collab., M.Derrick et al., Z. Phys. C73 (1996) 73; \\
H1 Collab., S. Aid et al., Nucl. Phys. B463 (1996) 3; \\
ZEUS Collab., M.Derrick et al., Z. Phys. C73 (1997) 253.
\bibitem{ref:regge} 
see e.g. P.D.B.~Collins, {\em Introduction to Regge
Theory and High Energy Physics}\\ 
(Cambridge University Press, 1977).
\bibitem{ref:vdm} 
J.J.~Sakurai, {\em Currents and Mesons} (University of Chicago Press,
  1969);\\
H.~Fraas and D.~Schildknecht, Nucl. Phys., B14 (1969)~543.
\bibitem{ref:dl} A.~Donnachie and P.V.~Landshoff, Phys. Lett. B296
(1992)~227.
\bibitem{ref:psi} ZEUS Collab., M.Derrick et al., Phys. Lett. B350
(1995)~120; \\
H1 Collab., S.~Aid et al., Nucl. Phys., B472 (1996)~3.
\bibitem{ref:zeuselpsi}
ZEUS Collab., J.~Breitweg et al., Z. Phys. C75 (1997)~215.
\bibitem{ref:disvm1} EMC Collab., J.J.~Aubert et al., Phys. Lett. B161
(1985)~203.
\bibitem{ref:nmc} NMC Collab., M.~Arneodo et al., Nucl. Phys. B429 (1994)~503.
\bibitem{ref:e665} E665 Collab., M.R.~Adams et al., Z.Phys. C74 (1997)~237.
\bibitem{ref:disvm2} ZEUS Collab., M. Derrick et al., Phys. Lett. B356
(1995)~601; \\
ZEUS Collab., M.~Derrick et al., Phys. Lett. B380 (1996)~220.
\bibitem{ref:h1dis} 
H1 Collab., S.~Aid et al., Nucl.Phys. B468 (1996)~3;\\
Erratum in preparation for Nucl.Phys. B.
\bibitem{ref:bfgms} 
S.J.~Brodsky et al., Phys. Rev. D50 (1994)~3134.
\bibitem{ref:ryskin} 
M.G.~Ryskin, Z. Phys. C57 (1993)~89.
\bibitem{ref:gluon}
ZEUS Collab., M.~Derrick et al., Phys. Lett. B345 (1995)~576;\\
H1 Collab., S.~Aid et al., Nucl. Phys., B470 (1996)~3. 
\bibitem{ref:ryslla}
M.G.~Ryskin, R.G.~Roberts, A.D.~Martin and E.M.~Levin, Z. Phys. C76 (1997)~231.
\bibitem{ref:cdm}
J.~Nemchik, N.N.~Nikolaev and B.G.~Zakharov, Phys. Lett.~B341 (1994)~228;\\
J.~Nemchik, N.N. Nikolaev, E.~Predazzi and B.G.~Zakharov,\\ Phys. Lett.~B374 (1996)~199;\\
J.~Nemchik, N.N.~Nikolaev, E.~Predazzi and B.G.~Zakharov, Z. Phys.~C75 (1997)~71;\\
J.~Nemchik et al., LANL Preprint HEP--PH/97-12-469 (1997).
\bibitem{ref:dl_vm} A.~Donnachie and P.V.~Landshoff, Phys. Lett. B348 (1995)~213;\\
J.R.~Cudell and I.~Royen, Phys. Lett. B397 (1997)~317.
\bibitem{ref:ginzburg} I.F.~Ginzburg and D.Yu.~Ivanov, Phys. Rev. D54
(1996)~5523;\\
D.Yu.~Ivanov, Phys. Rev. D53 (1996)~3564.
\bibitem{ref:collins}
J.C.~Collins, L.~Frankfurt and M.~Strikman, Phys. Rev. D56 (1997)~2982.
\bibitem{ref:CTEQ3} 
CTEQ Collab., H.L.~Lai et al., Phys. Rev. D51 (1995)~4763.
\bibitem{ref:fks}
L.~Frankfurt, W.~Koepf and M.~Strikman, Phys. Rev. D54 (1996)~3194.
\bibitem{ref:fs_dis95}
L.~Frankfurt and M.~Strikman, LANL Preprint HEP--PH/95-10-291 (1997).
\bibitem{ref:mrt} 
A.D.~Martin, M.G.~Ryskin and T.~Teubner, Phys. Rev. D55 (1997)~4329.
\bibitem{ref:zeus} 
The ZEUS Detector, Status Report, DESY (1993).
\bibitem{ref:cal} 
M.~Derrick et al., Nucl. Instrum. Methods A309 (1991)~77;\\
A.~Andresen et al., Nucl. Instrum. Methods A309 (1991)~101;\\
A.~Bernstein et al., Nucl. Instrum. Methods A336 (1993)~23.
\bibitem{ref:ctd} 
N.~Harnew et al., Nucl. Instrum. Methods A279 (1989)~290;\\
B.~Foster et al., Nucl. Phys. B, Proc-Suppl. B32 (1993)~181; \\
B.~Foster et al., Nucl. Instrum. Methods, A338 (1994)~254.
\bibitem{ref:lumi} 
D.~Kisielewska et al., DESY-HERA 85-25 (1985);\\
J.~Andruszk\'ow et al., DESY 92-066 (1992).
\bibitem{ref:hand} 
L.N.~Hand, Phys. Rev. 129 (1963)~1834.
\bibitem{ref:angle} 
K.~Schilling and G.~Wolf, Nucl. Phys. B61 (1973)~381.
\bibitem{ref:schc} 
J.~Ballam et al., Phys. Rev. D5 (1972)~545;\\
P.~Joos  et al., Nucl. Phys.  B113  (1976)~53;\\
F.J.~Gilman, J.~Pumplin, A.~Schwimmer and L.~Stodolsky, Phys.~Lett.~B31 (1970)~387.
\bibitem{ref:trigger} 
W.H.~Smith et al., Nucl. Instrum. Methods A355 (1995)~278.
\bibitem{ref:neural} 
H.~Abramowicz, A.~Caldwell and R.~Sinkus, Nucl. Instrum. Methods A365 (1995)~508; \\
R. Sinkus, Ph.D. thesis, Hamburg University (1996). 
\bibitem{ref:teresa} 
T.~Monteiro, Ph.D. thesis, Hamburg University (1998).
\bibitem{ref:beier} 
H.~Beier, Ph.D. thesis, Hamburg University (1997).
\bibitem{ref:jetset} 
T.~Sj\"ostrand, Comp. Phys. Commun. 39 (1986)~347;\\
T.~Sj\"ostrand and M. Bengtsson, Comp. Phys. Commun. 43 (1987)~367.
\bibitem{ref:heracles} 
A.~Kwiatkowski, H.~Spiesberger and H.-J.~Moehring,
in {\em Proceedings of the Workshop on Physics at HERA, Volume~III},
edited by W.~Buchm\"uller and G.~Ingelman (DESY, Hamburg, Germany, 1991),
p.~1294.
\bibitem{ref:muchor} 
K.~Muchorowski, Ph.D. thesis, Warsaw University (1998).
\bibitem{ref:dipsi} 
M.~Arneodo, L.~Lamberti and M.G.~Ryskin, Comp. Phys. Commun. {100} (1997)~195.
\bibitem{ref:mrsaprime} 
A.D.~Martin, W.J.~Stirling and R.G.~Roberts, Phys. Lett. B354 (1995)~155.
\bibitem{ref:epsoft} 
M. Kasprzak, Ph.D. thesis, Warsaw University (1996).
\bibitem{ref:herwig}
G.~Marchesini et al., Comp. Phys. Commun. 67 (1992)~465;\\
B.R. Webber, 
in {\em Proceedings of the Workshop on Physics at HERA, Volume~III},
edited by W.~Buchm\"uller and G.~Ingelman (DESY, Hamburg, Germany, 1991),
p.~1354;\\ 
L. Stanco, ibid., p.~1363.
\bibitem{ref:pythia} 
T.~Sj\"{o}strand, Z. Phys. C42 (1989)~301.
\bibitem{ref:flab} 
CDF Collab., F.~Abe et al., Phys. Rev. D50 (1994)~5535. 
\bibitem{ref:fact} 
For reviews, see e.g.\\
G.~Alberi and G.~Goggi, Phys. Rep. {74} (1981)~1;\\
K.~Goulianos, Phys. Rep. {101} (1983)~169;\\
N.P.~Zotov and V.A.~Tsarev, Sov. Phys. Uspekhi 31 (1988)~119;\\
G.~Giacomelli, Int.~J.~Mod.~Phys.~A, Volume 5 (1990)~223.
\bibitem{ref:rhopdissZEUS} 
ZEUS Collab., J.~Breitweg et al., E. Phys. J. C2 (1998)~247.
\bibitem{ref:rhopdissH1}  
H1 Collab., C.~Adloff et al., Z. Phys. C75 (1997)~607.
\bibitem{ref:ming} S.~Wang, Ph.D. thesis, University of Iowa (1998).
\bibitem{ref:lpair}  
J.A.M.~Vermaseren, Nucl. Phys. B229 (1983)~347; 
S.P.~Baranov, O.~D\"unger, H.~Shooshtari, and J.A.M.~Vermaseren, 
in {\em Proceedings of the Workshop on Physics at HERA, Volume~III}, 
edited by W.~Buchm\"uller and G.~Ingelman (DESY, Hamburg, Germany, 1991), 
p.~1478.
\bibitem{ref:soeding} 
P.~S\"{o}ding, Phys. Lett. 19 (1966)~702.
\bibitem{ref:e401} 
E401 Collab., M.~Binkley et al., Phys. Rev. Lett. 48 (1982)~73.
\bibitem{ref:e516} 
E516 Collab.,  B.H.~Denby et al., Phys. Rev. Lett. 52 (1984)~795.
\bibitem{ref:zeus94nlo}
ZEUS Collab., M.~Derrick et al., Z. Phys. C72 (1996)~399.
\bibitem{ref:mrsr2}
A.D.~Martin, R.G.~Roberts and W.J.~Stirling, Phys. Lett. B387 (1996)~419.
\bibitem{ref:grv94}
M.~Gl\"uck, E.~Reya and A.~Vogt, Z. Phys. C67 (1995)~433.
\end{thebibliography}
\end{document}